\documentclass{mmg2}
\usepackage{amsmath,amsfonts,amssymb}
\usepackage{url}
\DeclareMathOperator{\sgn}{sgn}

\begin{document}

\markboth{K. Wawrzyniak and W. Wislicki}
{Multi-market minority game: breaking the symmetry of choice}

%
%

\title{MULTI-MARKET MINORITY GAME:\\ 
BREAKING THE SYMMETRY OF CHOICE
}

\author{KAROL WAWRZYNIAK}

\address{Interdisciplinary Centre for Mathematical and Computational Modelling, University of Warsaw, Pawi\'nskiego 5A, PL-02-106 Warszawa, Poland\\
K.Wawrzyniak@icm.edu.pl}

\author{WOJCIECH WISLICKI}

\address{A. So\l tan Institute for Nuclear Studies, Ho\.za 69, PL-00-681 Warszawa, Poland\\
wislicki@fuw.edu.pl}

\maketitle

\begin{history}
\end{history}

\begin{abstract}
Generalization of the minority game to more than one market is considered.
At each time step every agent chooses one of its strategies and acts on the market related to this strategy.
If the payoff function allows for strong fluctuation of utility then market occupancies become inhomogeneous with preference given to this market where the fluctuation occured first.
There exists a critical size of agent population above which agents on bigger market behave collectively.
In this regime there always exists a history of decisions for which all agents on a bigger market react identically.
\end{abstract}


\section{Introduction}

Minority game (MG) is a model of adaptive behavior in multi-agent systems where being in minority is profitable \cite{challet_1}.
Each agent can individually adapt to variable state of the system by updating its choice algorithms, called strategies.
This adaptation can be done using data from the sequence of recent winning decisions.
Strategies are ranked every time by evaluating their utilities, dependent on the integrated population actions, such that the best strategy can be chosen.

The MG was originally formulated \cite{arthur_challet,challet_2} for one public sequence $\mu$ of winning minority decisions.
There was also one binary choice set, available for each agent, and actions were defined on the same set of events.
Overall {\it demand}, being a sum of individual actions, was calculated for the whole set of agents.
Using analogy to bargain on a stock exchange, where actions are interpreted as {\it sell} or {\it buy}, we call such game a {\it single-market minority game} (SMG).
\begin{figure}[h]
\begin{center}
\includegraphics[scale=.35]{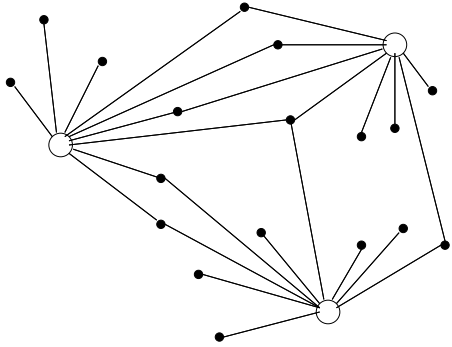}
\end{center}
\vspace*{8pt}
\caption{\label{fig1} Representation of MMG where each agent (black point) is linked to the market (white point) where it acts.}
\end{figure}

Observing real financial markets one immediately finds that large numbers of investors use to act concurrently on many markets, usually different in size, deciding not only on their sell/buy actions but, first of all, choosing the market.
This naturally motivates to generalize the original minority game in order to account for such multi-market agent activity and call it the {\it multi-market minority game} (MMG).
The MMG can be visualized as in Fig.~\ref{fig1}, where the open symbols refer to markets, the black ones to agents and link between an agent and a market represents enablement of the agent to play on this market.
If the agent actually acts on the market the corresponding link is called {\it active}.

In this paper, among other things, we are particularly interested in distribution of market occupancies and its time evolution.
In many network systems it is observed \cite{scale_free} that a small number of nodes absorbs majority of active links.
We can see the similar phenomenon in the MMG where significant differences between numbers of agents acting on different markets arise in course of the game.
Further in the paper we call this phenomenon the {\it breakdown of the symmetry of choice}.

\section{Formal definition of the game}

The MMG consists of $N$ agents and $K$ markets.
At each time step $t$, the $n$-th agent $(n=1,\ldots,N)$ chooses a preferable market $k_n(t)\in\{1,\ldots,K\}$ and takes action $a_n(t)\in\{-1,1\}$ on this market.
For each market $k$ an aggregated demand is defined
\begin{eqnarray}
A_k(t)=\sum_{n=1}^{N_k(t)}a_n(t),
\label{eq1}
\end{eqnarray}
where $N_k(t)$ is the number of agents who chose market $k$ at time $t$.
The $A_k(t)$ is thus the difference between numbers of agents who choose the $+1$ and $-1$ actions.
Agents do not know each other's actions but $A_k(t)$ is known to all agents on the market.
For each market $k$ the minority action is determined
\begin{eqnarray}
a_k^\ast(t)=-\sgn A_k(t).
\label{eq2}
\end{eqnarray}
Each agent's memory is limited to $m$ most recent winning decisions.

Each agent has the same number $s\ge 2$ of devices, called strategies, per market, altogether $K\cdot s$ strategies per agent.
Strategies are used to predict the next minority action $a_k^\ast(t+1)$.
The strategy of the $n$-th agent, $\alpha_n$, is a function mapping the sequence $\mu$ of last $m$ winning decisions to this agent's action $a_n$.
Since for each market there is $P=2^m$ possible realizations of $\mu_k$, there is $2^P$ possible strategies per market.
At the beginning of the game each agent randomly draws $s$ strategies for each of $K$ markets, according to a given distribution function $\rho(n,k):N\times K\rightarrow \Delta_n^k$, where $\Delta_n^k$ is a set consisting of $s$ strategies.
For clarity, we give the flow chart of the game in Appendix A, where the processes of generation of the memory pools and taking decisions by agents are better illustrated.

Each strategy $\alpha_n$, belonging to any of sets $\Delta_n^k$, is given a real-valued function $U_{\alpha_n}$ which quantifies the utility of the strategy: the more preferable strategy, the higher utility it has.
Strategies with higher utilities are more likely chosen by agents.

There are various choice policies.
In the popular {\it greedy policy} each agent selects the strategy of the highest utility
\begin{eqnarray}
\alpha_n(t)=\arg \max_{\alpha_n \in \{\Delta_n^1,...,\Delta_n^k\}} U_{\alpha_n}(t).
\label{eq3}
\end{eqnarray}
The strategy (\ref{eq3}) used by the agent is called the {\it active strategy}, in contrast to {\it passive strategies} unused at given moment.
At given time an agent acts only on one market related to the active strategy.
However, every time it evaluates all its strategies, active and passive ones.
Each strategy $\alpha_n$ is given the {\it payoff} related to its action $a_{\alpha_n}$
\begin{eqnarray}
R_{\alpha_n}(t)=-a_{\alpha_n}(t)\,g[A(t)],
\label{eq4}
\end{eqnarray}
where $g$ is an odd function, e.g. $g(x)=\sgn(x)$ \cite{challet_2}.
Other choices are $g(x)=x/N$ or $g(x)=x$, the latter used in the present work.
The learning process corresponds to updating the utility for each strategy
\begin{eqnarray}
U_{\alpha_n}(t+1)=U_{\alpha_n}(t)+R_{\alpha_n}(t)
\label{eq5}
\end{eqnarray}
such that each agent knows how good its strategies are.

\section{Network structure}

\begin{figure}[h]
\begin{center}
\begin{tabular}{cc}
\includegraphics[scale=.25]{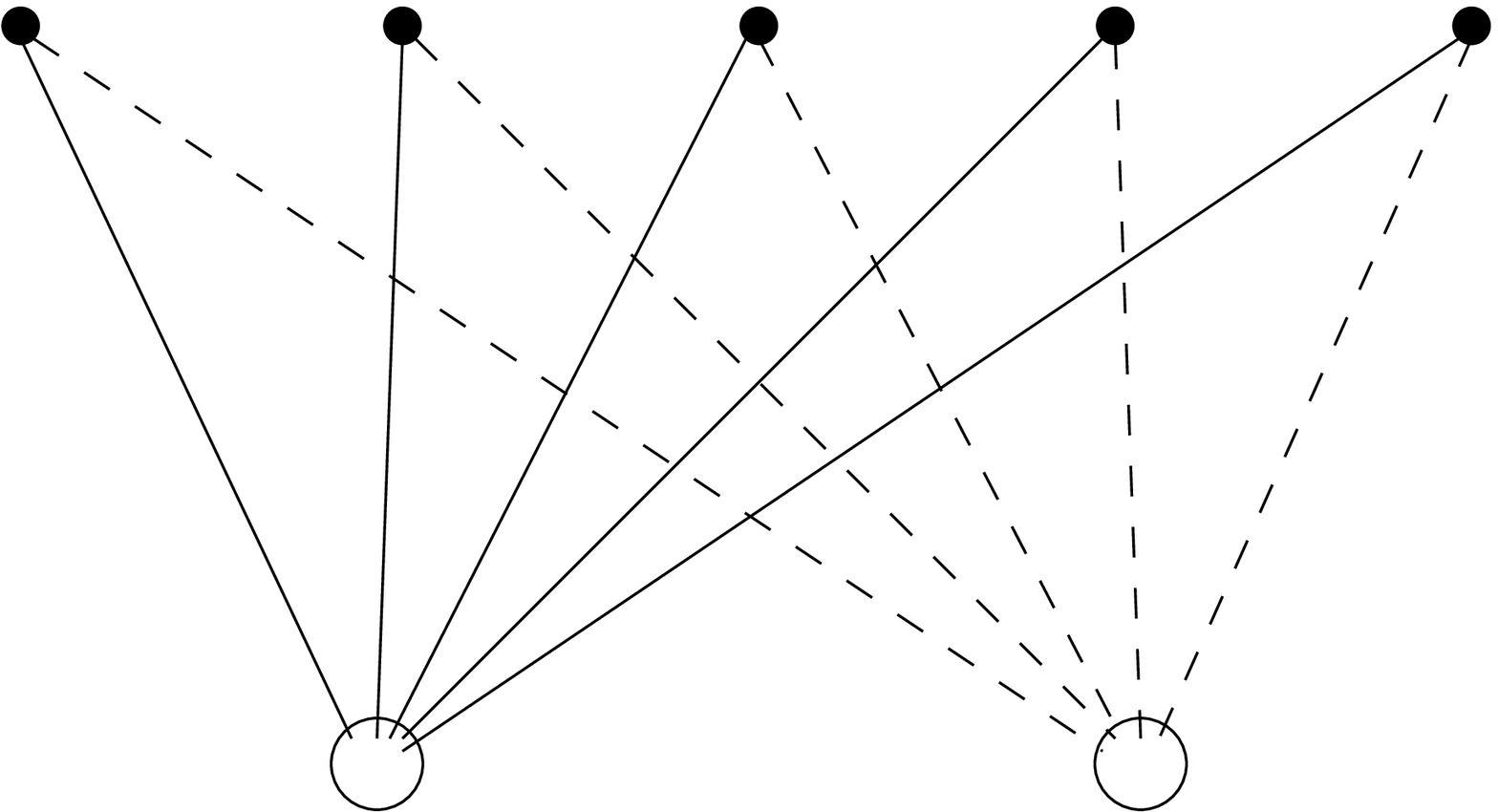} & \hspace{1cm} \includegraphics[scale=.25]{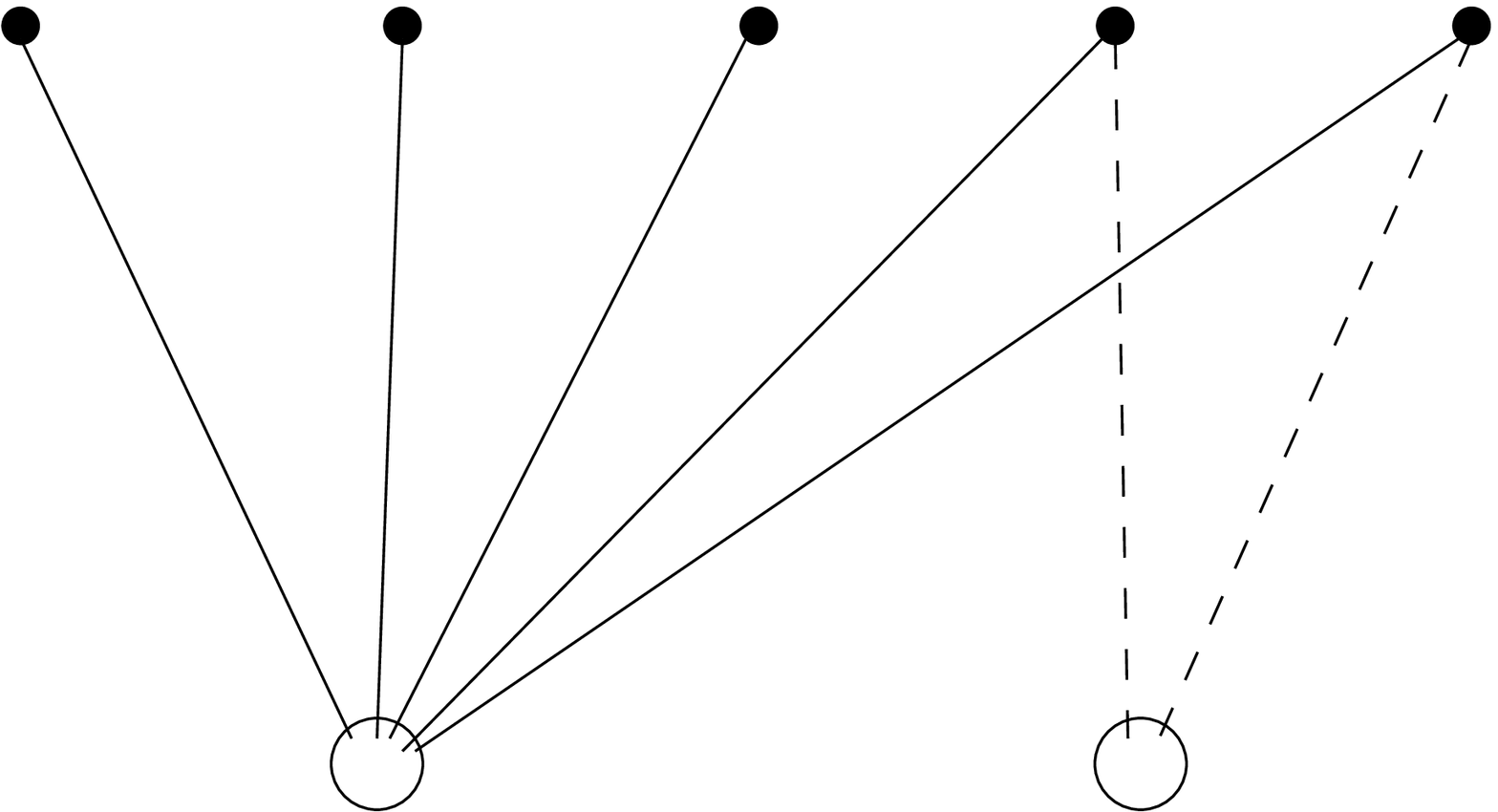}
\end{tabular}
\end{center}
\vspace*{8pt}
\caption{\label{fig2} The regular (left) and irregular (right) market structures. Full lines correspond to agent (black points) links to the first market (white points) and dashed ones to the second market.}
\end{figure}

We distinguish two types of markets: the {\it regular} where all $N$ agents can choose any market, and {\it irregular} one where $N_1$ agents can play only on one market and remaining $N_2=N-N_1$ agents can choose both (cf. Fig.~\ref{fig2}).

At any time, agents connected to more than one market have to choose only one of them to act.
Sets of strategies for markets and lengths of memories $m$ do not change during the game.
These two network structures are sufficient to demonstrate the phenomenon of the asymmetry of choice and its universality.
In particular, we observe that the phenomenon does not depend on the structure and is present for both topologies.
Irregular topology is more general and better reflects reality where numbers of agents playing on different markets may not be the same.

In this paper the MMG on two markets is studied in detail but we discuss also briefly the three-market case.

\section{Relation to other models}

The MMG was first discussed by the authors of ref. \cite{platkowski_1} and developed subsequently \cite{platkowski_2}.
The authors study in detail the rules used by each agent when switching between markets, and how the agent's performance depends on these rules.
Performance of agents is measured by the {\it volatility}, i.e. the variance of the aggregate demand normalized to the number of agents, for various schemes. 
It is found that the volatilities qualitatively behave similarly for all considered schemes.
However, the overall levels of volatility differ for different switching schemes.
In particular, the volatility in the {\it virtual scores scheme} where the highest-utility strategy is chosen, is relatively higher than in the other schemes in the cooperation mode \cite{platkowski_2}.
This switching scheme was used in the present paper.
Summarizing, our model coincides with the model of ref. \cite{platkowski_2} with the virtual scores scheme.
They exhibit the following features:
\begin{enumerate}
\item Individuals can choose a market from a set of markets of different size.
\item Each agent has at least two strategies per market.
\item The choice of the market and the action on it are preceded by the analysis of the self-generated information pattern $\mu$.
\item The strategy space is not reduced.
\end{enumerate}
As a result, such MMG represents a  direct extension to the classical MG. 
It is important to notice that the analysis in ref. \cite{platkowski_2} was performed in the cooperation regime of the game. 
Therefore the authors could not observe the phenomenon of breaking the symmetry of choice which shows up in the herd regime and is found and explained in our research, as described below.

Other known extentions to the classical MG formulate the multi-choice \cite{chan} or the multi-asset MGs with one \cite{bianconi} or more \cite{martino} strategies per asset.
The authors of ref.~\cite{bianconi} also observe the breakdown of the symmetry of choice but using different kind of model. 
They assume that public information pattern $\mu$ is chosen randomly and independently, and  with uniform probability, from the artificially created pool of information patterns for each asset.
A distinct feature of that MMG is an artificially introduced asymmetry of sizes of $\mu$ pools for each asset. 
This means that their MMG cannot use the history of the real past winning decisions because in such a case one could generate a history $\mu$ not included in the reduced pool.
Consequently, the history $\mu$ has to be chosen randomly from the reduced sets for each asset and for each time step.
The dynamical character of the game is then limited only to the strategy score. 
In our opinion, this change distracts the model from the original idea of the MG, whereas our model is a more straightforward extension of the genuine MG.
This modification, viz. the artificial asymmetry of the $\mu$ pools, is responsible for the market split observed in ref. \cite{bianconi}.
Therefore the symmetry breaking observed in ref. \cite{bianconi} is not a spontaneous one, contrary to our model where the sizes of information pools are equal.
To be more specific, we recall the notation used by the authors of ref.~\cite{bianconi} who define two memory pools $\mu_{\pm}$ and ratios $\alpha_\pm$ of their sizes to the number of agents
\begin{eqnarray}
\mu_{\pm} & \in & \{1,\ldots,P_{\pm}\}, \nonumber \\
\alpha_\pm & = & P_\pm/N,
\label{eq41}
\end{eqnarray}
where $P_{\pm}$ stand for sizes of these pools. These sizes are not necessarily maximal, i.e. they may be smaller than $2^m$.
This naturally limits the choice set and violates the symmetry of choice in case $\alpha_+\ne\alpha_-$ just because of unequal sizes of $\mu_+$ and $\mu_-$.
Contrary to that, in our model there is always $\alpha_+=\alpha_-$ and $\mu_\pm$ are maximal, i.e. $P_+=P_-=2^m$.
In this case the authors of ref. \cite{bianconi} do not observe any asymmetry but we do.
Our present paper aims to explain why in certain circumstances markets behave asymmetrically despite $\alpha_+=\alpha_-$.


\section{Asymmetry of choice}

\subsection{Regular markets}

\begin{figure*}[h]
\begin{tabular}{cc}
\includegraphics[scale=.24]{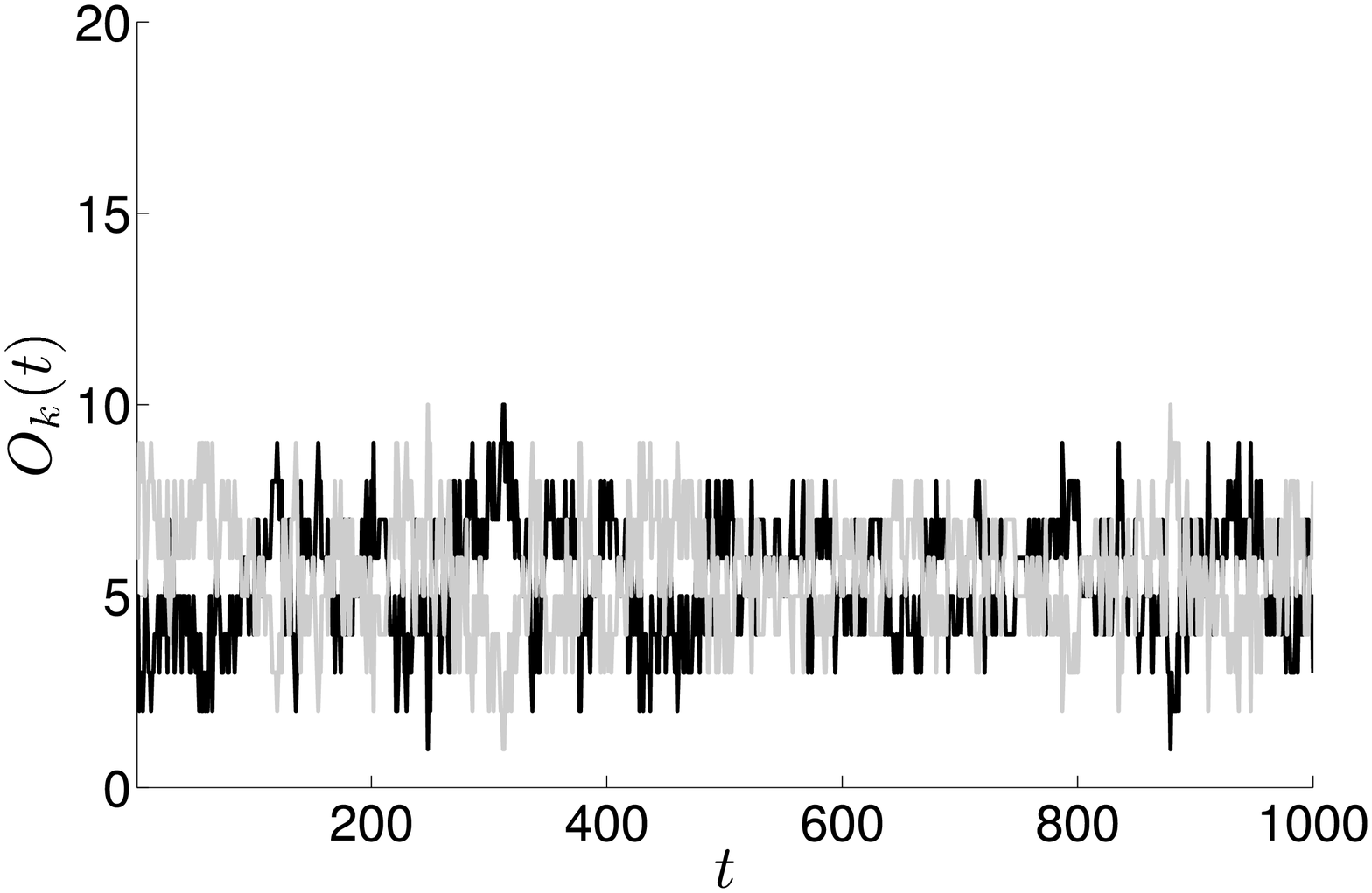} & \includegraphics[scale=.24]{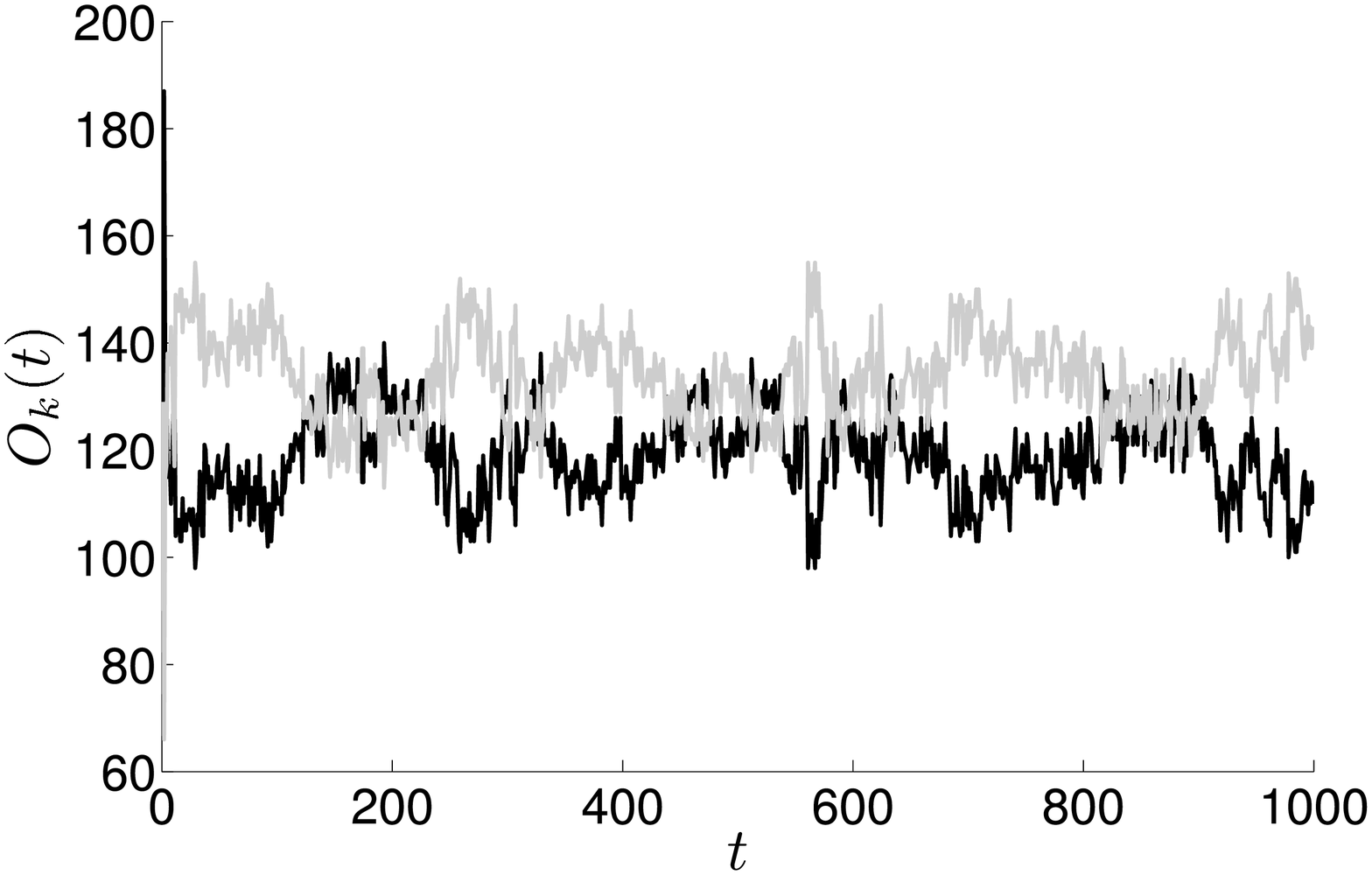} \\
\vspace*{8pt}
\includegraphics[scale=.24]{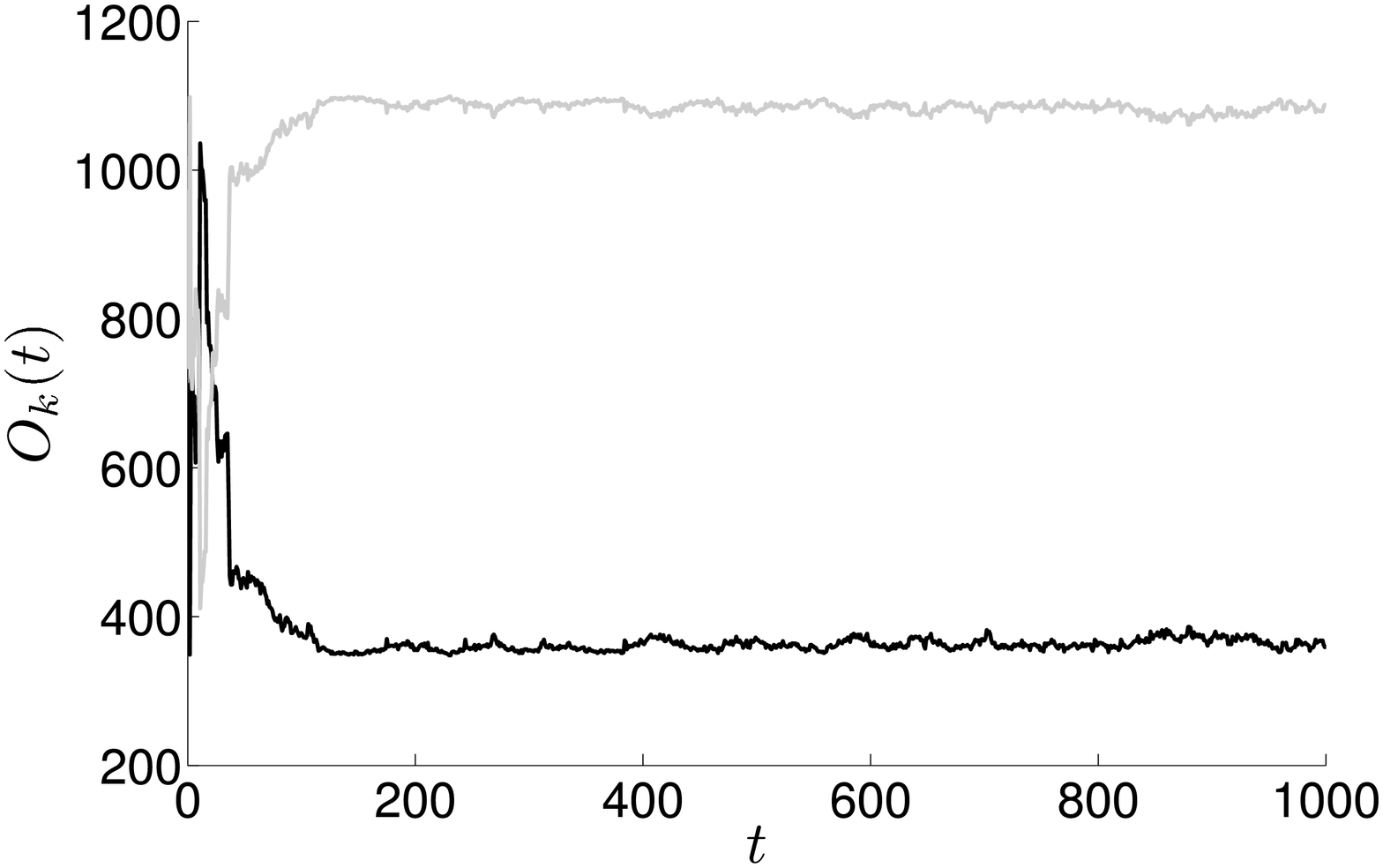} &
\end{tabular}
\vspace*{8pt}
\caption{\label{fig3} Time evolution of the market occupancy $O_k(t)$ (black lines correspond to $k=1$ and grey ones to $k=2$) in MMG for $m=5$ and for three agent populations $N=11$ (upper left), $253$ (upper right) and $1447$ (bottom). It is seen how the split of the market occupancies increases with $N$.}
\end{figure*}

\begin{figure*}[h]
\begin{tabular}{cc}
\includegraphics[scale=.24]{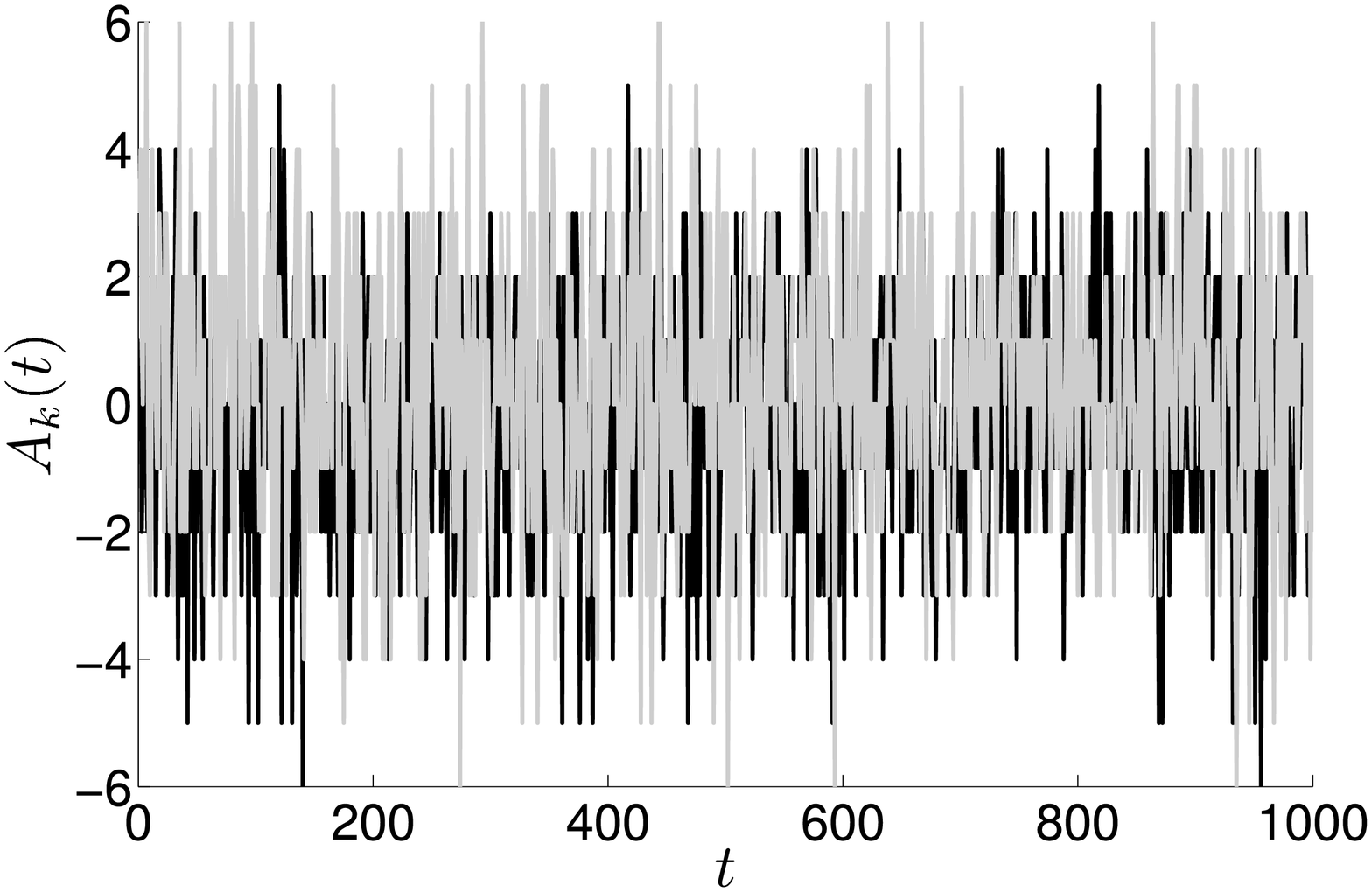} & \includegraphics[scale=.24]{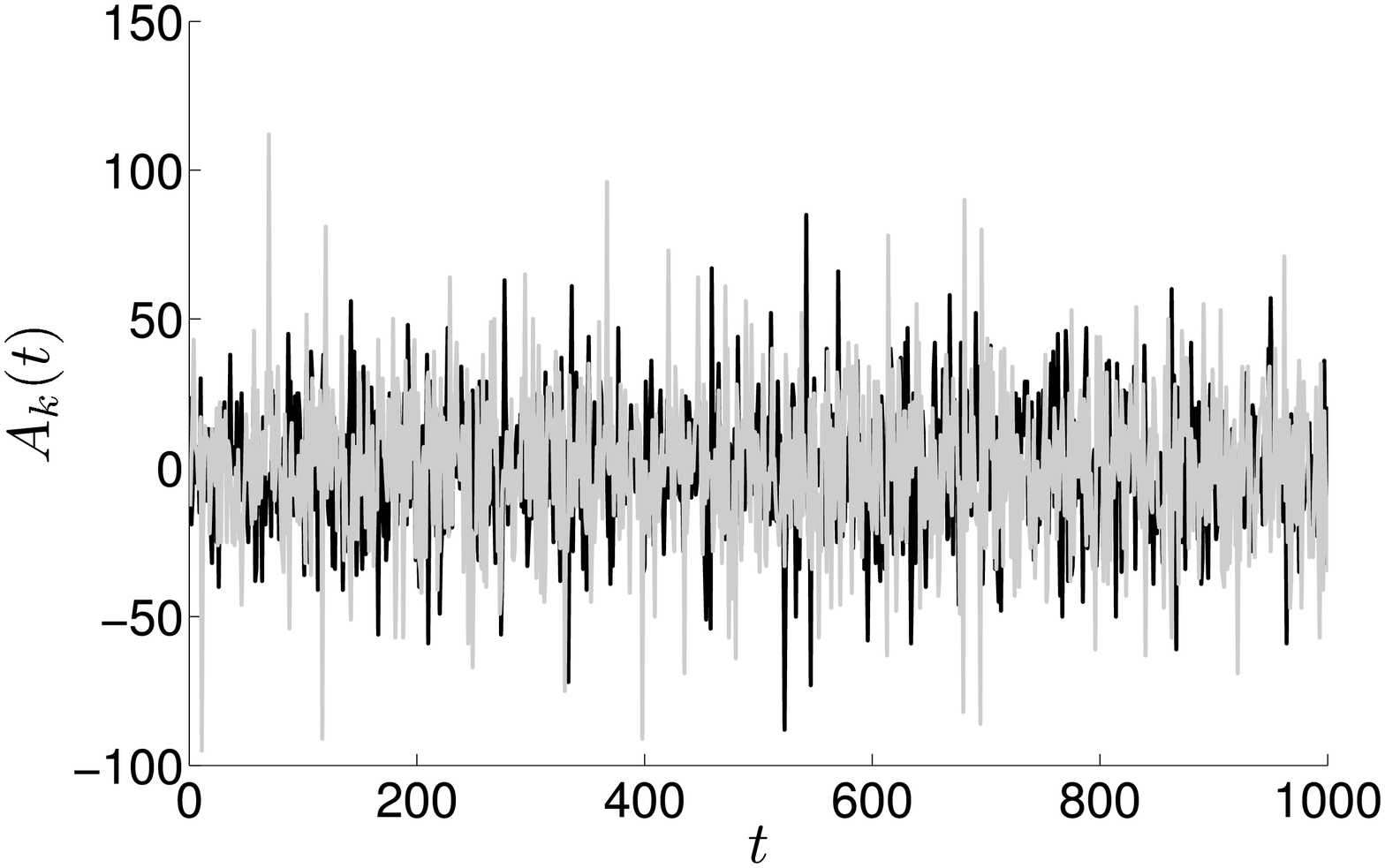} \\
\vspace*{8pt}
\includegraphics[scale=.24]{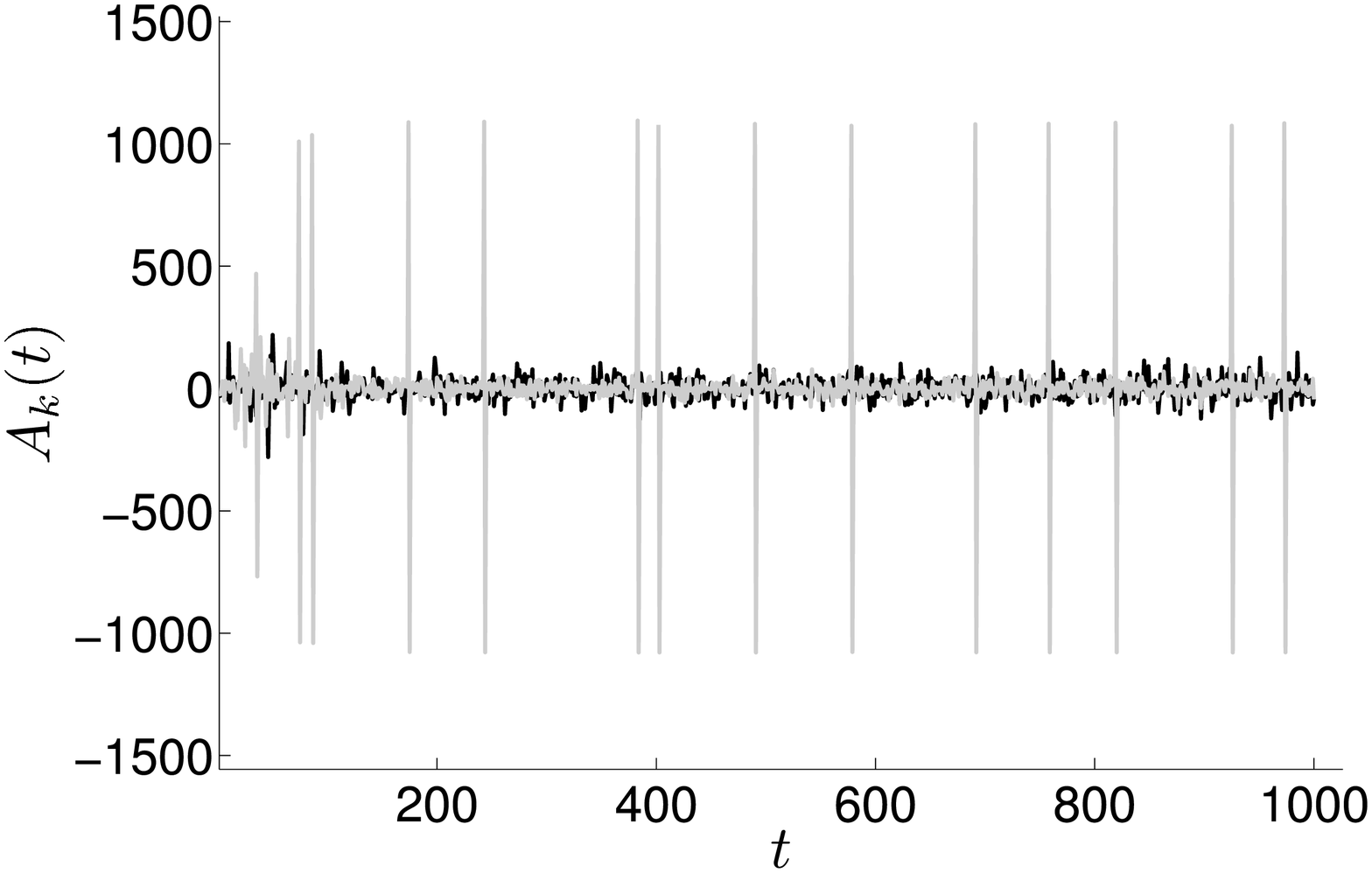}
\end{tabular}
\vspace*{8pt}
\caption{\label{fig4} Time evolution of the market demand $A_k(t)$ (black lines correspond to $k=1$ and grey ones to $k=2$) in the MMG for $m=5$ and for three agent populations $N=11$ (upper left), $253$ (upper right) and $1447$ (bottom). It is seen how the fluctuations of $A_2(t)$ become more distinct for large $N$.}
\end{figure*}

After performing numerical simulation for the regular markets we plot the $k$th market {\it occupancy} $O_k(t)$, defined as the number of agents acting on this market at given time, and the $k$th market demand $A_k(t)$, for three total agent populations $N=11$, $253$ and $1447$ (cf. Figs~\ref{fig3} and \ref{fig4}).

\noindent These numbers correspond to different modes of the game: the random ($N=11$ and $m=5$) and the herd ($N=253, 1447$ and $m=5$), according to the terminology used in ref.~\cite{savit} for the SMG.
Comparing behavior of $O_k(t)$ for different populations one observes that for large $N$ most agents prefer the same market to act (we call this market {\it big}) and one sees that $O_k(t)$ tends to stabilize fast.
There is no such effect for small $N$.
These observations are robust with respect to small variations of $N$.
Simulating ensemble of histories we checked that both markets are equally likely to become big.
Another effect, seen only on a big market, is an appearance of distinct peaks of $A_k(t)$ (Fig.~\ref{fig4}, right), comparable in size with the values of $O_k(t)$ for big market.
\begin{figure*}[h]
\begin{tabular}{cc}
\includegraphics[scale=.24]{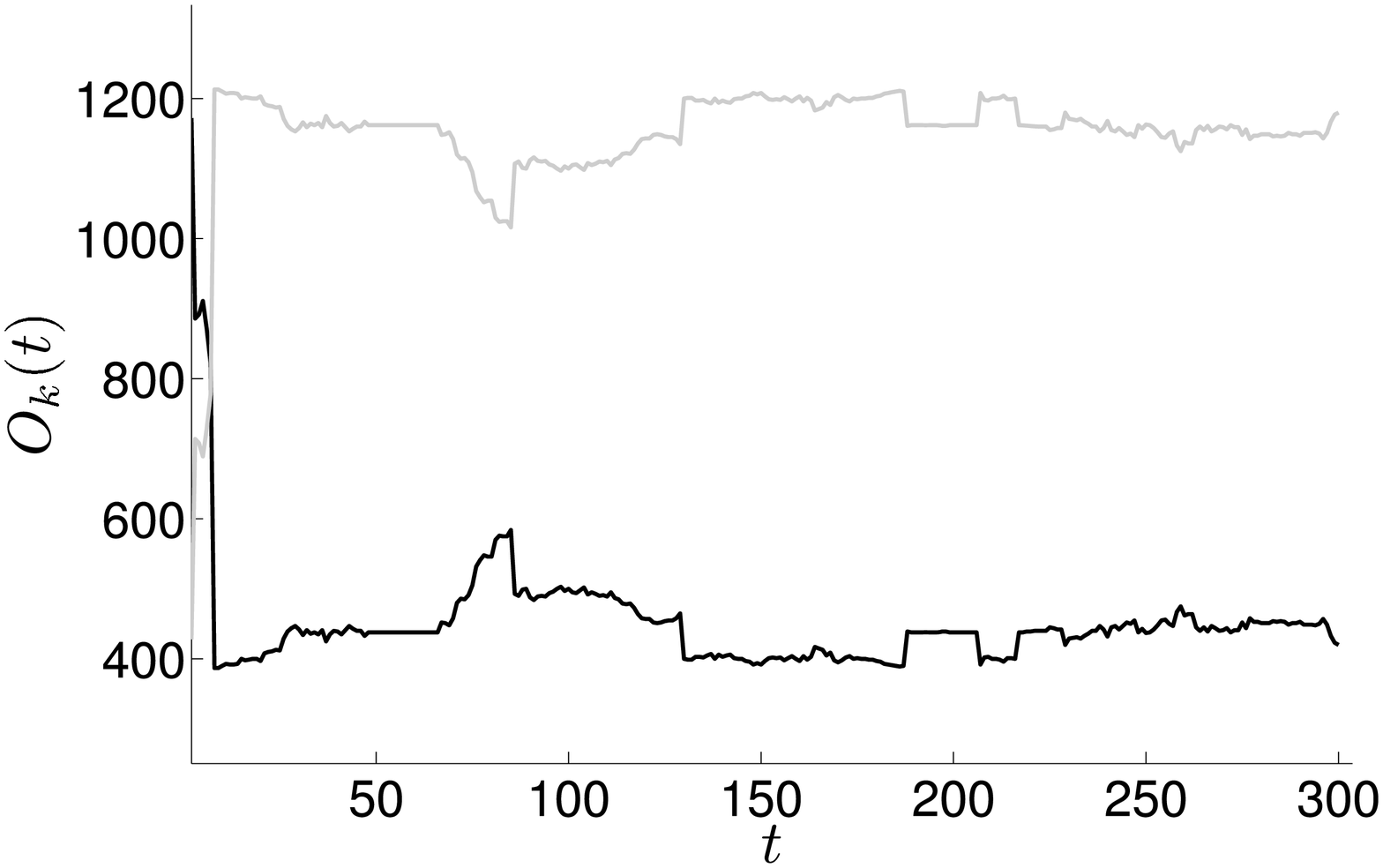} & \includegraphics[scale=.24]{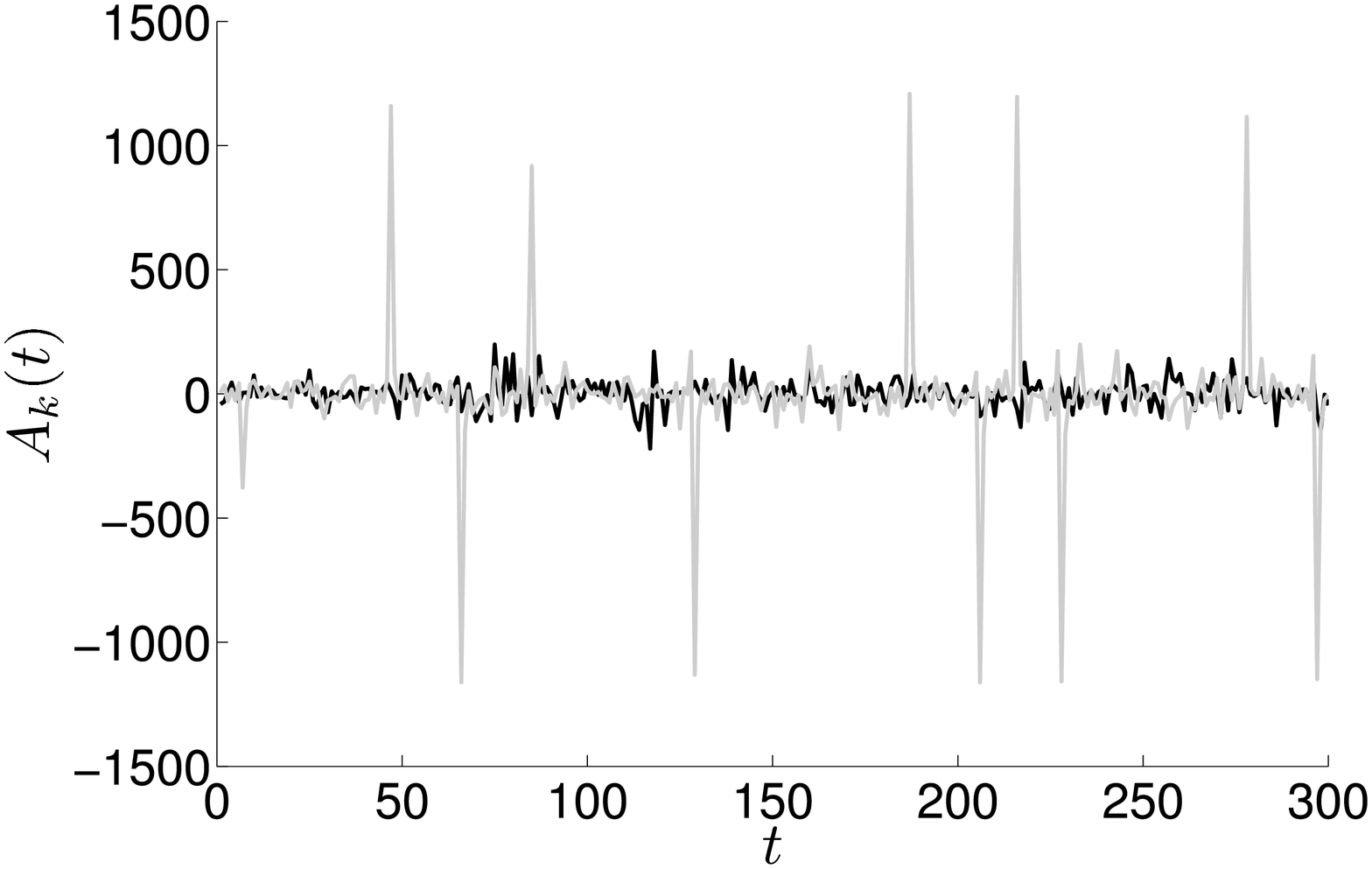} \\
\includegraphics[scale=.24]{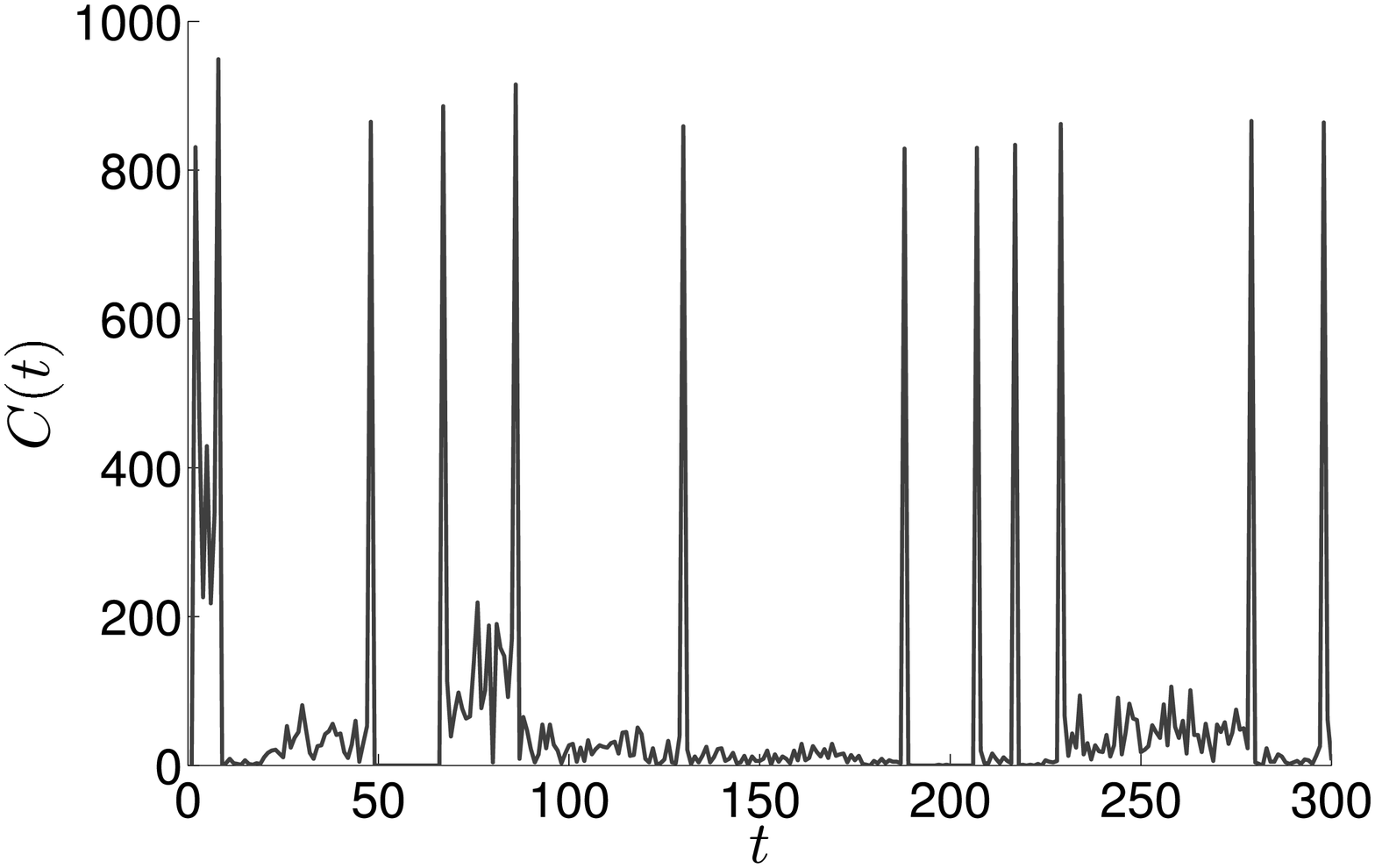}
\end{tabular}
\caption{\label{fig5}Time evolution of the market occupancy $O_k(t)$ (upper left), aggregated demand $A_k(t)$ (upper right) (black lines correspond to $k=1$ and grey ones to $k=2$) and number of agents changing prefered market $C(t)$ (bottom) in the MMG with $N=1600$, $s=2$ and $m=5$.}
\end{figure*}
\begin{figure*}[h]
\begin{tabular}{cc}
\includegraphics[scale=.24]{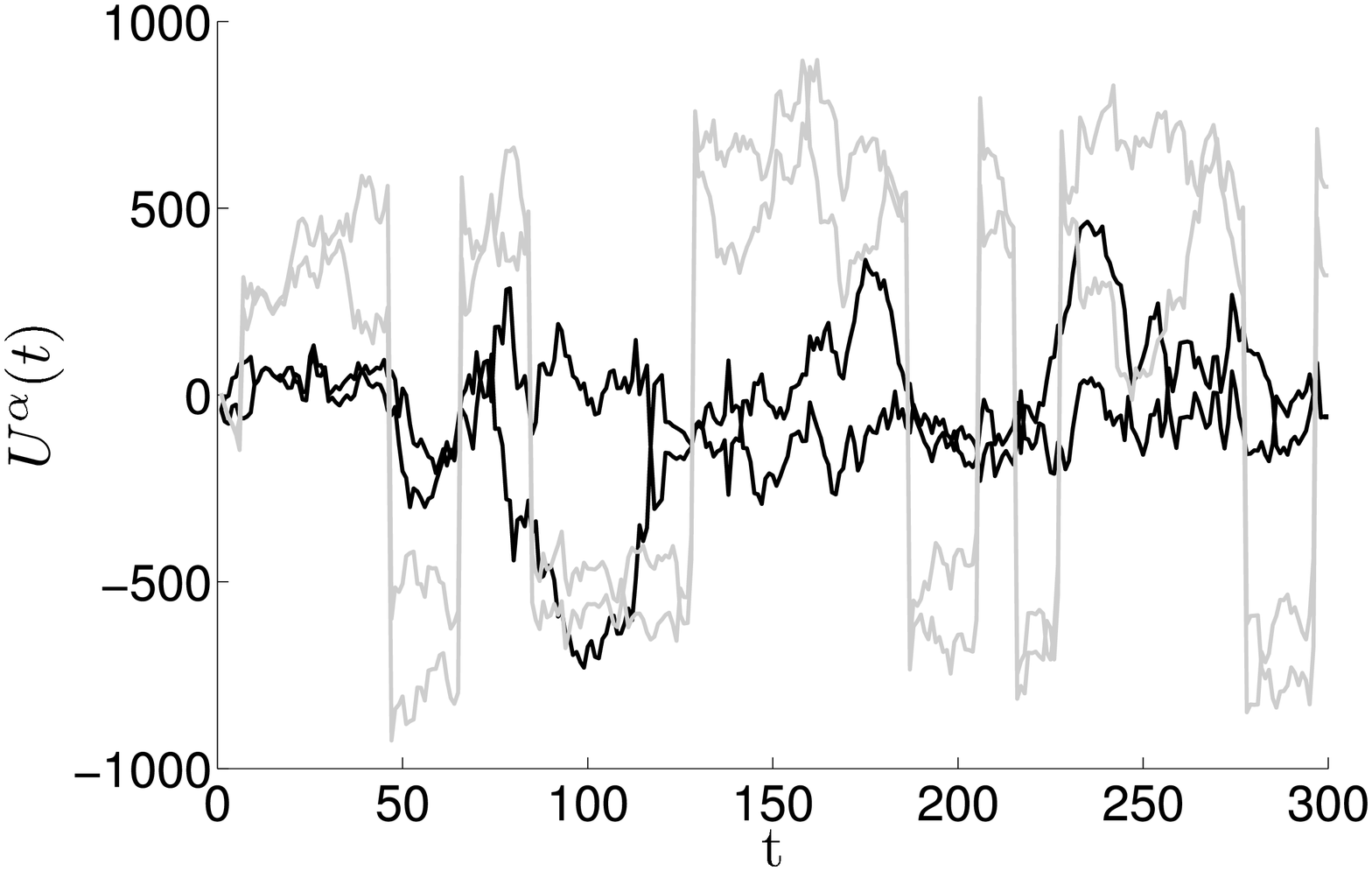} & \includegraphics[scale=.24]{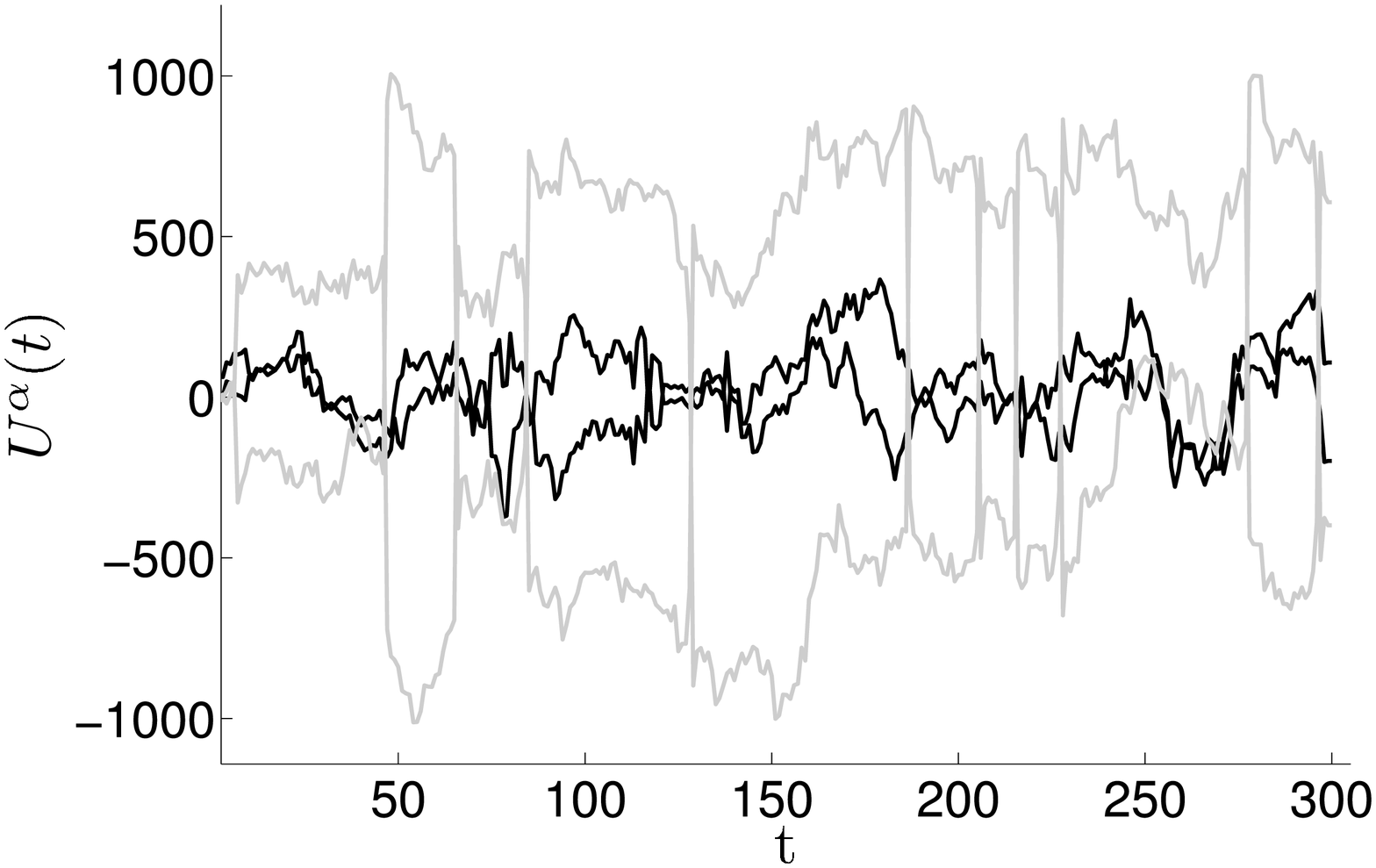} \\
\includegraphics[scale=.24]{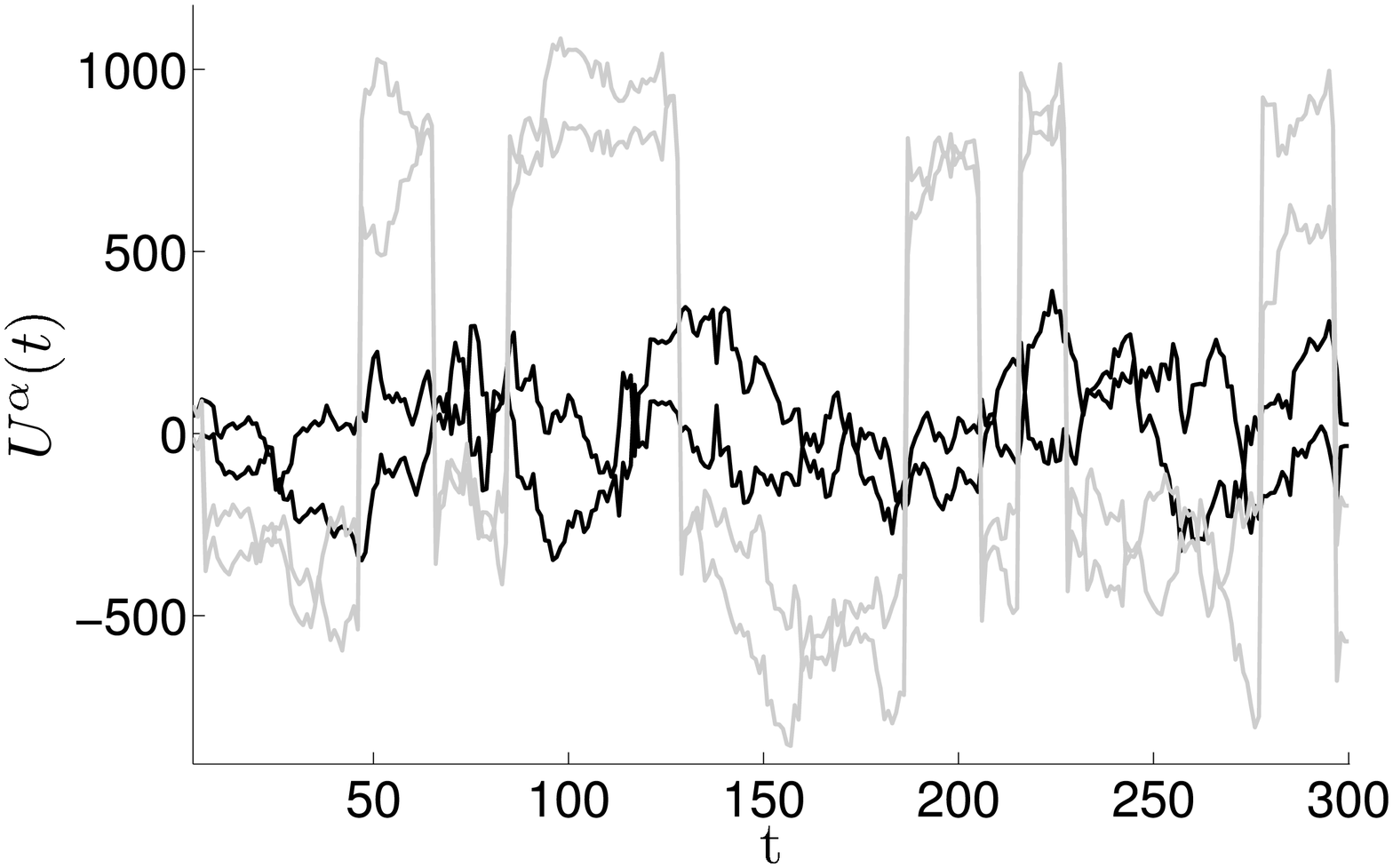}
\end{tabular}
\caption{\label{fig6} Time evolution of the utility for 4 strategies, for the MMG with $N=1600$, $s=2$ and $m=5$, for three representative agents: with two high-utility strategies on market 1 at the time $t_1+1=8$ (upper left), one high- and one low-utility strategy on market 1 at the time $t_1+1=8$ (upper right) and two low-utility strategies on market 1 at $t_1+1=8$ (bottom). Grey lines, thin and bold, correspond to 2 strategies on market 1 and black ones, thin and bold, on market 2.}
\end{figure*}

In order to study the effect further, consider the following regular game with large $N$:
\begin{enumerate}
\item For any given history $\mu$ of winning strategies, about half of strategies $\alpha$ suggest action $a=1$ and another half $a=-1$,
\item Utilities of strategies are uniformly distributed at initial time.
\end{enumerate}
For such defined MMG we perform 300 steps of evolution and in Fig.~\ref{fig5} we plot $O_k(t)$, $A_k(t)$ and the number of agents who change the market $C(t)$.
Fig.~\ref{fig6} presents utility evolution for three agents, each of them using two different strategies on two markets.

Depending on the sign of the payoff (\ref{eq4}) associated with the minority decision (\ref{eq2}), the strategy is called {\it good} ({\it bad}) for the positive (negative) payoff.
The probability of the output of any strategy on the $k$-th market is the same as for the minority decision on the $k$-th market and is equal to $1/2$.
Thus the probability that an agent has no good strategies is equal to $1/2^s$.
Subsequently, the probability of having at least one good strategy is equal to $1-1/2^s$.
For example, for $s=2$ the $75\,\%$ of the population, in the limit $N\rightarrow\infty$, has at least one good strategy on market 1.

Since the $A_k(t)$ fluctuates \cite{arthur_challet,challet_3}, in both the SMG and MMG, rapid change is also observed in the utility $U(t)$, according to eqs (\ref{eq4}) and (\ref{eq5}).
If $A_k(t_1)$ strongly fluctuates at time $t_1$, the utility of strategies related to market $k$ is strongly affected.
This can be seen in Fig.~\ref{fig5} where at $t=7$ high value of $A_2$ shows up for the first time.
Consequently, utilities of strategies on market 2 change proportionally to $A_2(7)$, as seen in Figs~\ref{fig6}.
At $t_1+1$, about $1-1/2^s$ of agents have at least one strategy with high $U(t_1+1)$ on the market where the fluctuation of $A$ occured first.
All these agents choose this market.
As seen in Fig.~\ref{fig5}, at $t=8$ about $75\,\%$ of the population has at least one strategy with high utility on market 2.
Only agents with two bad strategies stay with market 1.

Since large fluctuation $A_k(t_1)$ comes after some history $\mu_k^C=\mu_k(t_1)$, all agents which have at least one strategy suggesting the same output as the minority action $a_k^\ast(t_1)$ for this history tend to choose the bigger market in the next step.
The $\mu^C$ has a non-vanishing probability to reappear at $t_2>t_1$.
Consequently, all agents belonging to bigger market react on the same way and $A_k(t_2)$ for this market deviates maximally, i.e. $A_k(t_2)=O_k(t_2)$, and no agent on the big market belongs to the minority.
In Fig.~\ref{fig5}, the $\mu_2^C$ appears again at $t=47$ and $A_2(47)=O_2(47)\simeq 1200$.

At $t_2$, all strategies with high $U(t_2)$ on bigger market fail and get penalty $-A_k(t_2)$, whereas strategies with low $U(t_2)$ are rewarded with $A_k(t_2)$.
As seen in Figs~\ref{fig6}, a quarter of the population with two high-utility strategies on bigger market now has two strategies with low utility.
This part decides to change the market.
Another quarter of the population with two low-utility strategies now has two strategies with high $U(t)$ and switches from the smaller market to the bigger one.
Remaining half of the population has one low- and one high-utility strategy.
This group remains in the similar situation and stays on a bigger market, although the utilities of strategies are swapped.
Generally, when $\mu_k^C$ appears then about half of agents changes market but $O_k(t)$ remains constant.

\begin{figure}[h]
\begin{tabular}{cc}
\includegraphics[scale=.22]{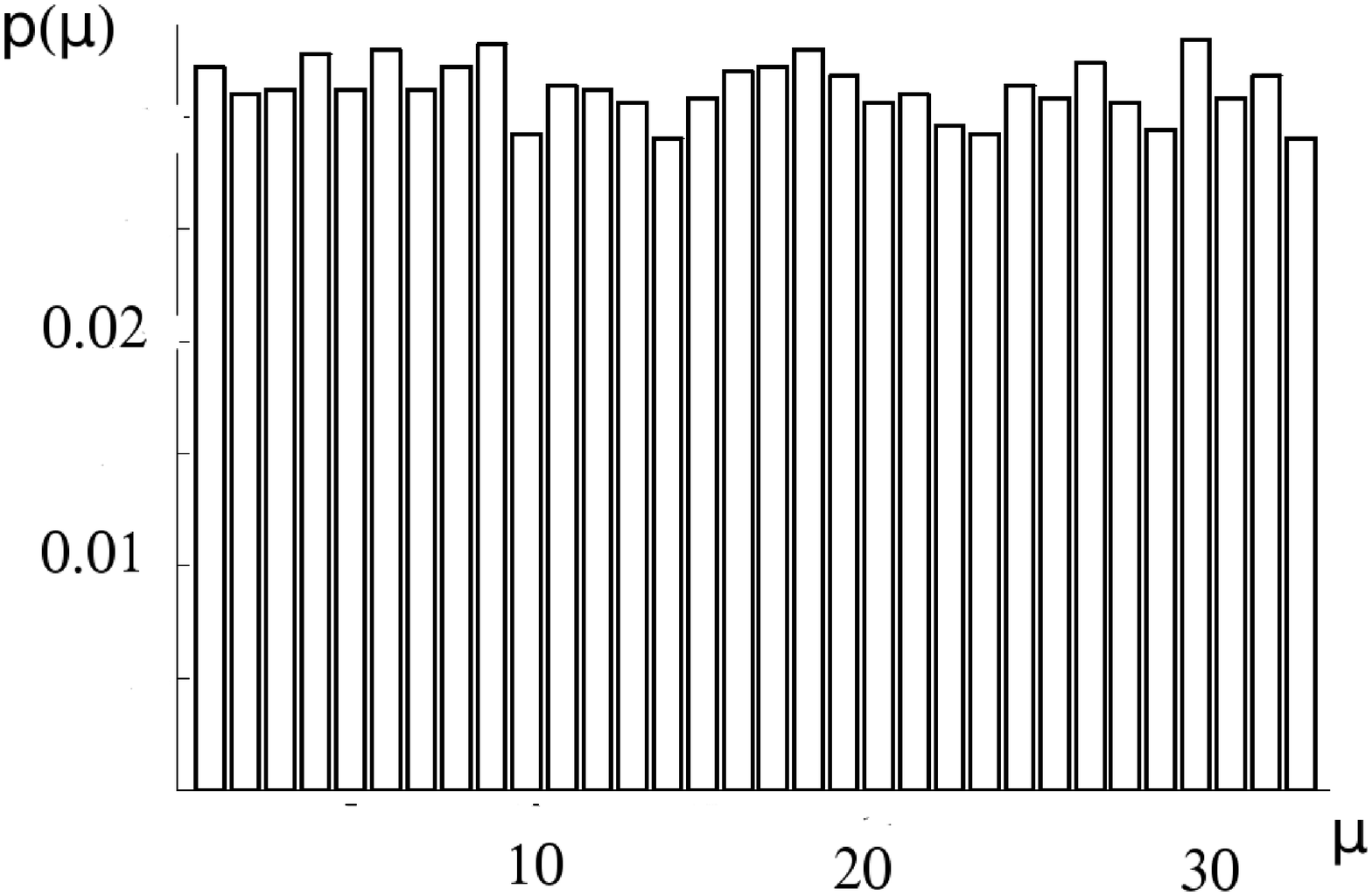} & \includegraphics[scale=.22]{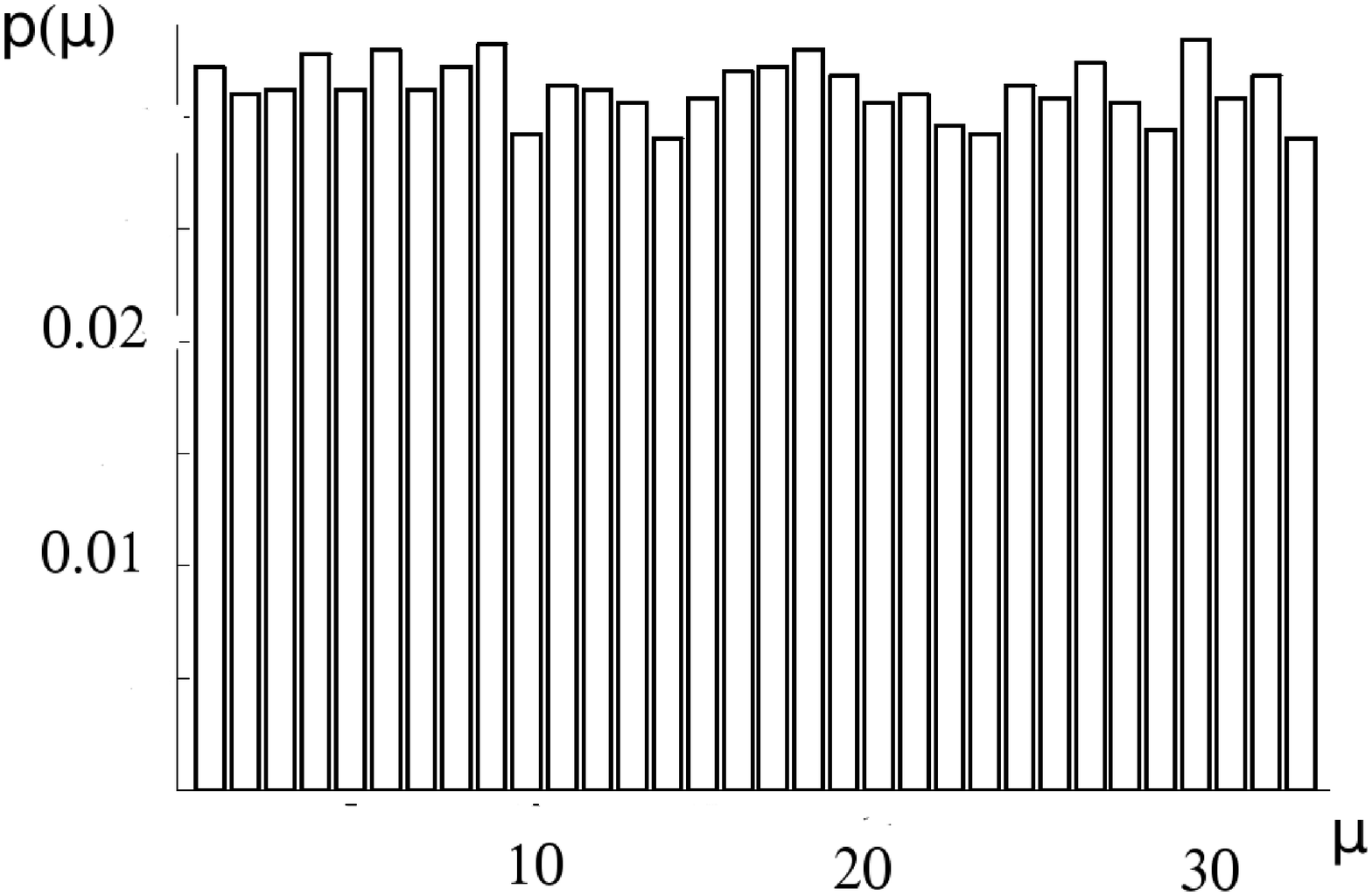} \\
\includegraphics[scale=.13]{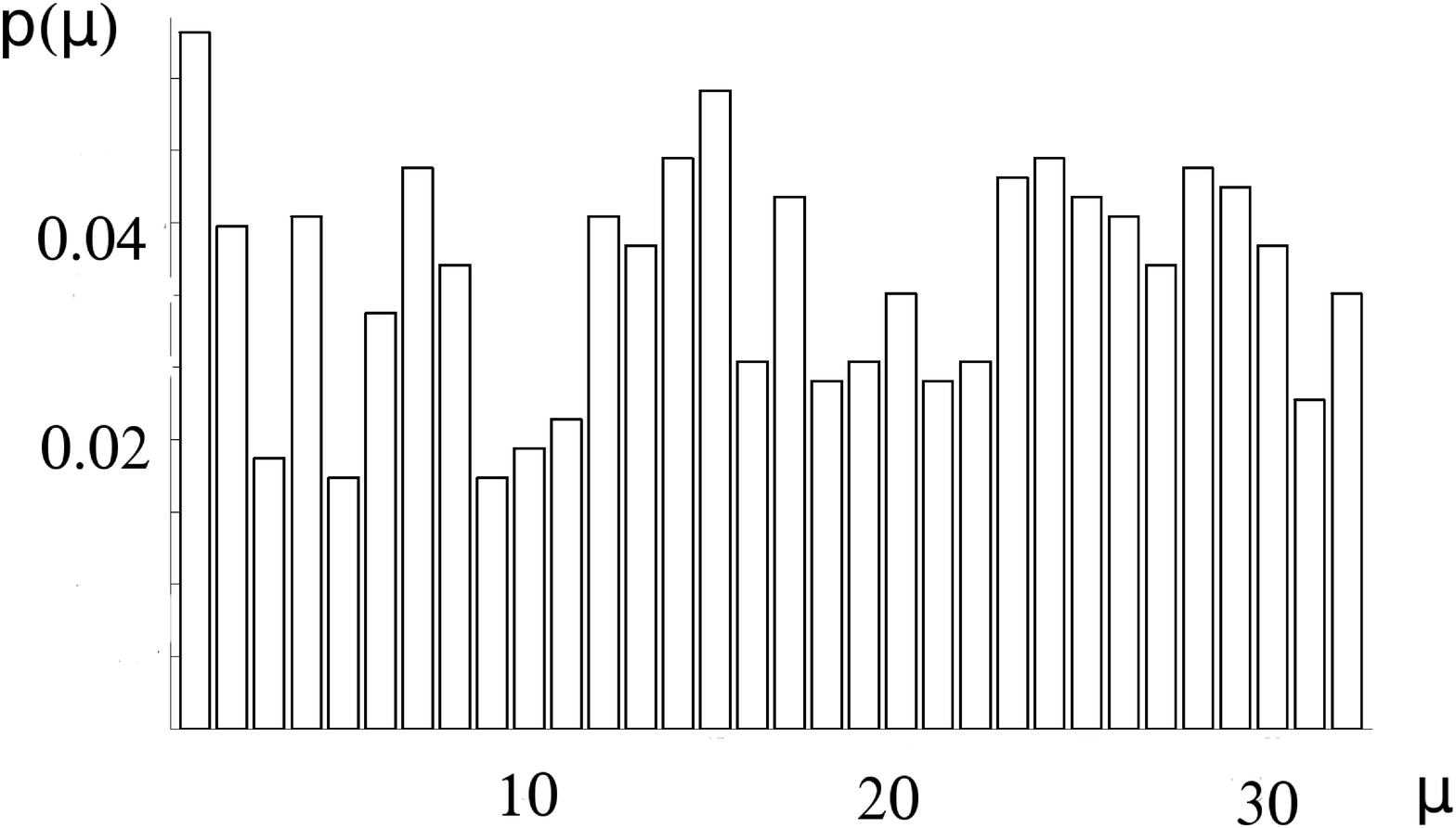} & \includegraphics[scale=.13]{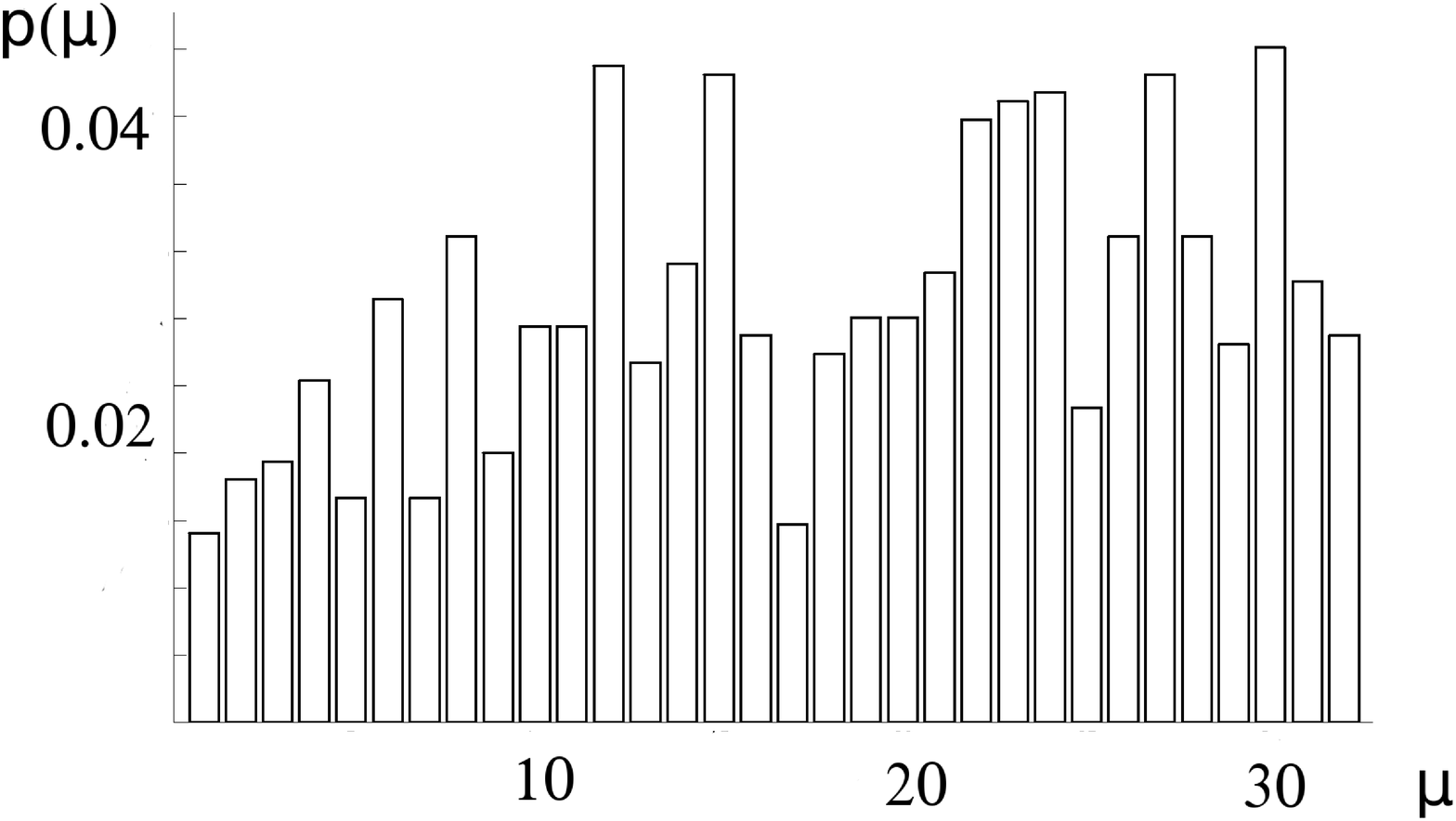}
\end{tabular}
\caption{\label{fig6_0} Probability density $p(\mu)$ for simulated histories $\mu$ for MMG with $N=1447$ and $m=5$, for the small (upper left) and large (upper right) markets, and the same probabilities for $N=11$, for the small (lower left) and large (lower right) markets. The $p(\mu)$ was estimated from 5,000 time steps for each game.}
\end{figure}

Another important characteristics of fluctuation dynamics is given by the fluctuation frequency.
In the asymptotic regime, the frequency of large fluctuations of $A_k$ on bigger market is equal to
\begin{eqnarray}
\langle\nu\rangle \sim 1/2^m, \quad\quad\mbox{for}\quad N\rightarrow\infty.
\label{eq5_p}
\end{eqnarray}
Each history appears, on average, once per $2^m$ times and a large fluctuation of $A_k$ occurs always after the history $\mu^C$.
We checked numerically that simulated $\langle\nu\rangle$ converges to the asymptotic value (\ref{eq5_p}) when simulation time and $N$ become large.
We also checked that, indeed, simulated histories $\mu$ are distributed uniformly with probabilities $p(\mu)$ around $1/2^m=1/32\;\;(m=5)$, as seen in Figs~\ref{fig6_0} (upper).
Contrary to that, for small $N$ some histories may be significantly more frequent than the others, as seen in Figs~\ref{fig6_0} (lower).
In such a case, the pool of strategies for the whole population is not large enough, their choice space is restricted and some histories occur more often.
Similar inhomogeneity of $p(\mu)$ was observed in case of the SMG, in somewhat different representation, by the authors of ref.~\cite{savit}.

\begin{figure}[h]
\begin{center}
\includegraphics[scale=.3]{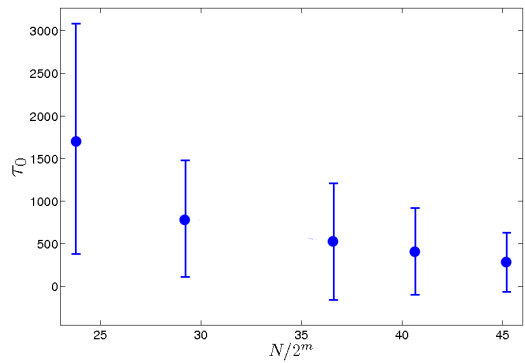}
\end{center}
\caption{\label{fig6_1} Relaxation time $\tau_0$ of the market occupancy $O_k(t)$ versus $N/2^m$ for $m=5$. The $\tau_0$ is a mean from ten games and error bars correspond to one standard deviation.}
\end{figure}

Interesting is the dynamics of the MMG when approaching its stationary regime.
In order to illustrate this, we examined the relaxation time of market occupancies.
As seen in Fig.~\ref{fig3} (right), for large population the stationarity of the $O_k(t)$ at levels $N/2^s$ and $N(1-1/2^s)$ is reached after finite time.
This time is related to the first occurence of strong fluctuation of $A_k(t)$, as discussed.
Defining $\tau_0$ as the time when $O_k(t)$ enters the $\pm 5\%$ belt around $N/2^s$ or $N(1-1/2^s)$ and stays inside it forever, we present in Fig.~\ref{fig6_1} how $\tau_0$ depends on the often used variable $Q=N/2^m$, where $m=5$.
Decreasing $\tau_0(N/2^m)$ is consistent with our observation that for large agent population the probability of strong fluctuation of demand $A_k$ is more likely.
Note however that $\tau_0$ is meaningless for too small $N$ when market occupancies do not stabilize at their asymptotic levels.

\begin{figure}[h]
\begin{tabular}{cc}
\includegraphics[scale=.23]{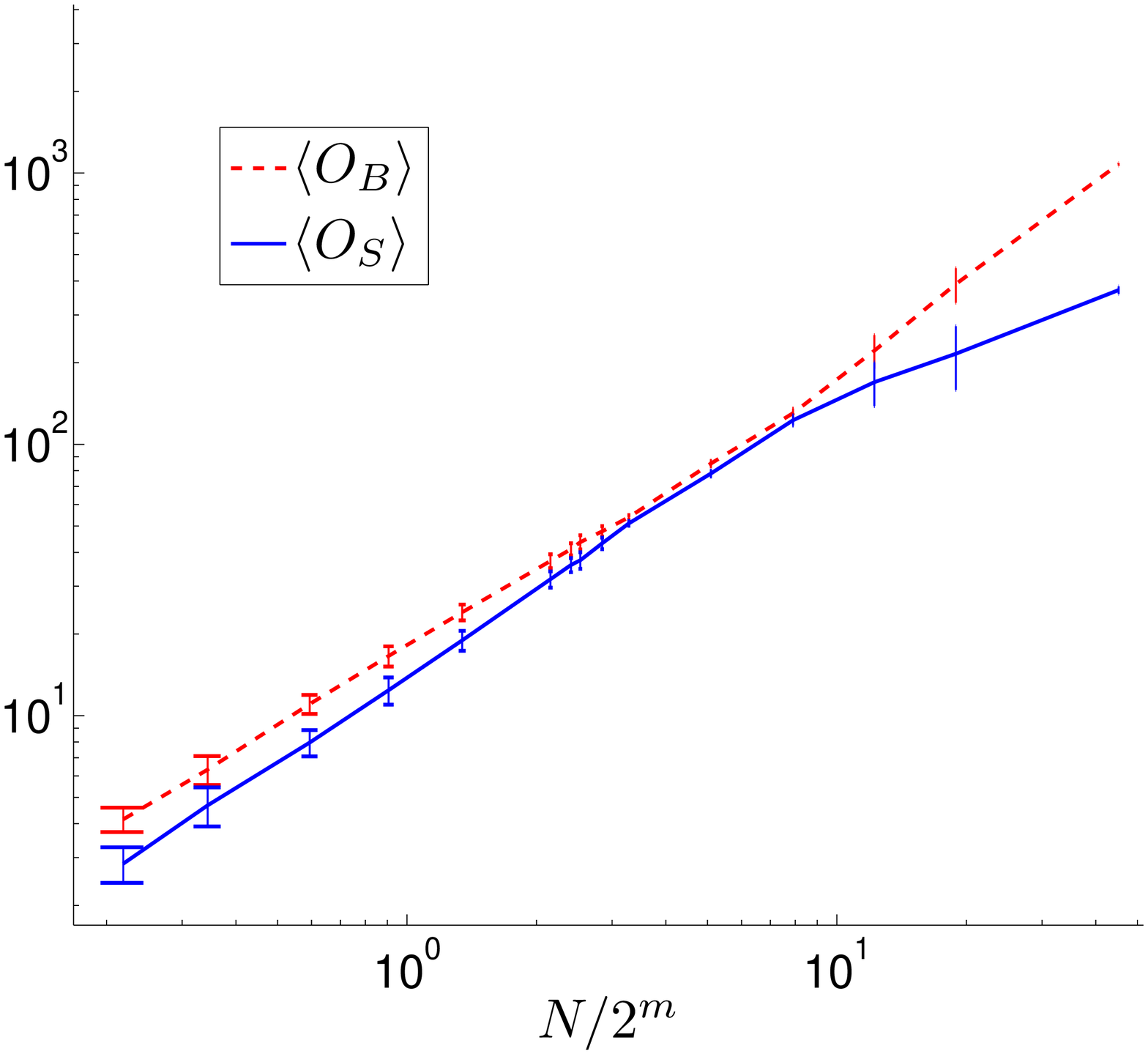} & \includegraphics[scale=.23]{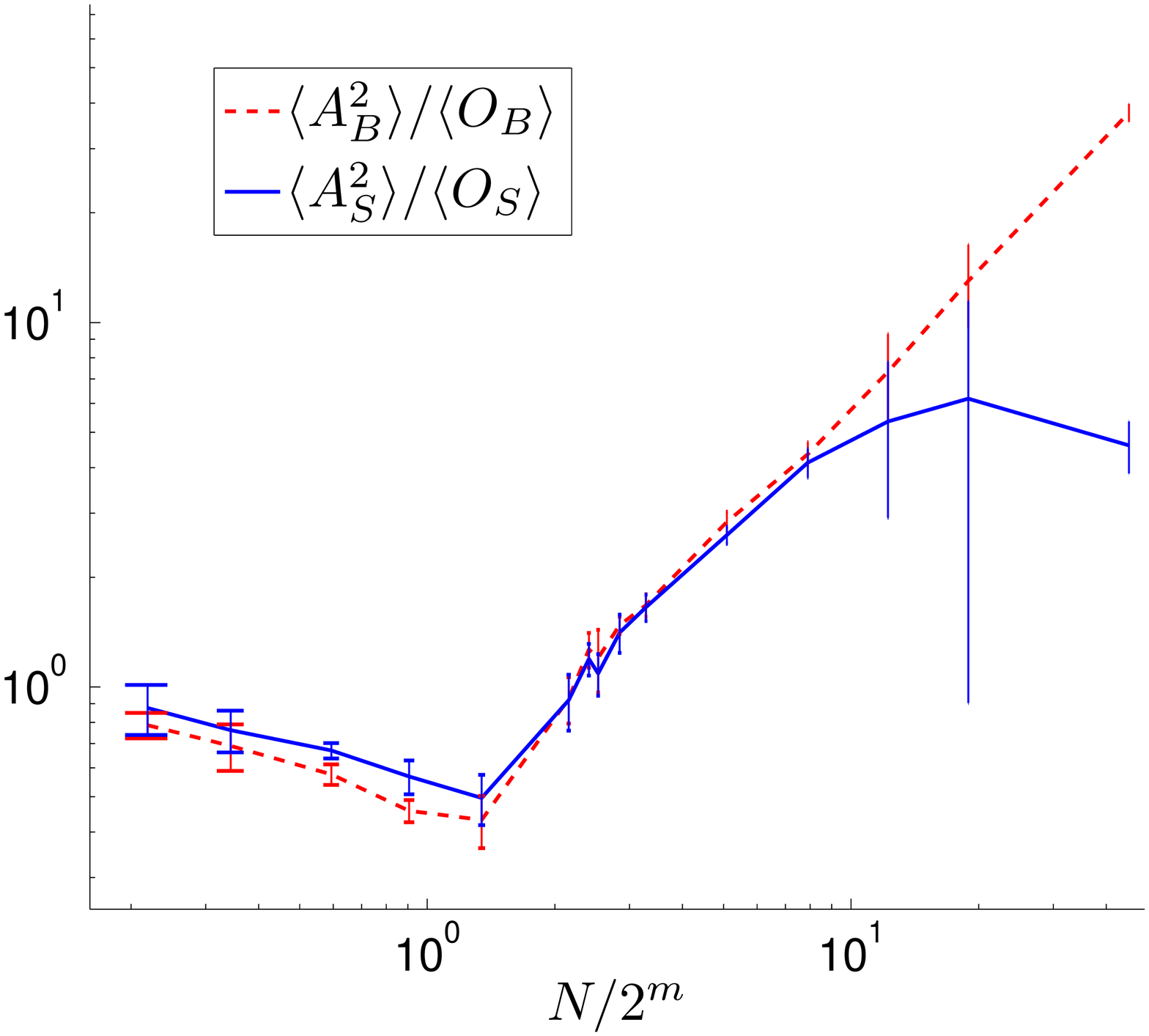}
\end{tabular}
\caption{\label{fig7} Mean occupancies of markets $\langle O\rangle$ (left) and variances of {\it per capita} demands $\langle A^2\rangle/\langle O\rangle$ (right) as functions of $Q=N/2^m$, where $m=5$ and $N$ is variable. Solid lines correspond to the small market and dotted lines to the big market. The markets have regular structure. Error bars correspond to one standard deviation and curves are drawn to guide ones eye.}
\end{figure}

Another interesting feature of the MMG is seen when the asymmetry of choice is examined as a function of an overall population size $N$.
Figs~\ref{fig7} show dependence of the mean occupancies and demand variances {\it per capita} on $Q=N/2^m$.
This dependence is shown separately for the big and small markets.
At $Q_c \simeq 8$ the qualitative change of behaviour is observed.
Below $Q_c$ both markets are about equally populated and have similar variance of $A_k$, whereas above $Q_c$ one of the markets becomes visibly bigger and has larger variance.
Below $Q_c$ the population is too small to generate fluctuation in $A_k$ sufficient to significantly affect utilities and split the agent set into two unequal groups.
\begin{figure}[h]
\begin{tabular}{cc}
\hspace{-.5cm} \includegraphics[scale=.24]{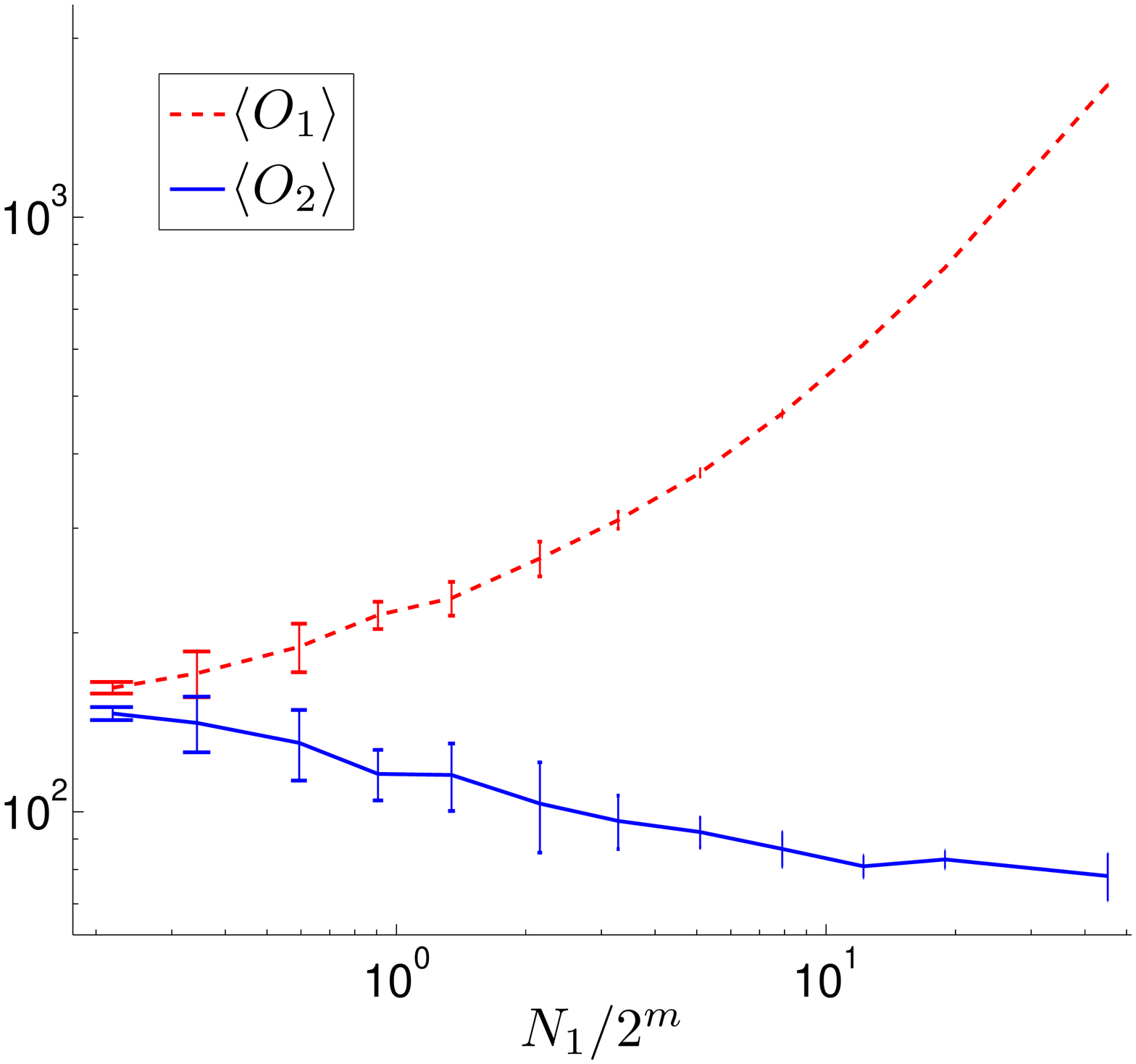} & \includegraphics[scale=.24]{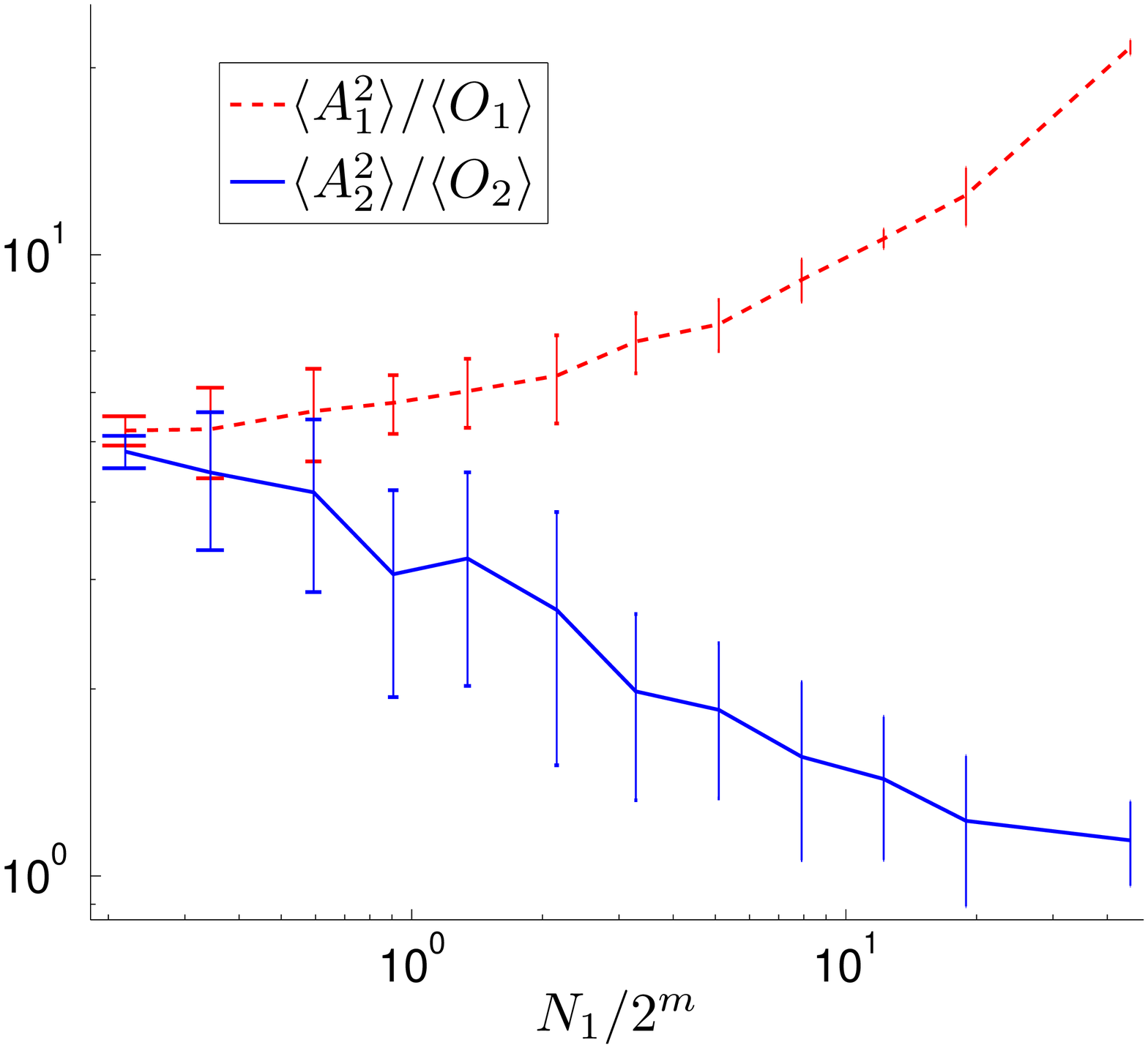}
\end{tabular}
\caption{\label{fig8} Mean occupancies of markets $\langle O\rangle$ (left) and variances of {\it per capita} demands $\langle A^2\rangle/\langle O\rangle$ (right) as functions of $Q=N_1/2^m$, for variable $N_1$, fixed $N_2=301$ and $m=5$. Solid lines correspond to  market 1 and dotted lines to market 2. The markets have irregular structure. Error bars correspond to one standard deviation and curves are drawn to guide ones eye.}
\end{figure}
Our {\it per capita} demand fluctuations $\langle A_k^2\rangle/\langle O_k\rangle$ calculated for the MMG differ from that in the SMG where the quantity $\langle A^2\rangle/N$ is considered.
Our definition of the MMG assures that simplifying MMG to SMG, which is tantamount to switching to the classical MG, one gets identical games.

The SMG is known to exhibit three modes of behavior: the random, cooperation and herd \cite{savit}.
In the MMG, in addition, the fourth mode can be identified.
We call this behaviour {\it herd asymmetric}, to be distinguished from the {\it herd symmetric} where no market split is observed.
In this terminology the values of $Q\in\{0.34, 7.9, 45.2\}$ chosen for our simulations correspond to the random, herd symmetric and herd asymmetric modes.
Asymmetry appears at the transition point $Q_c$ where fluctuations {\it per capita} for the small market become lower than for the big market. Using economic terminology, the small market becomes more efficient than the big one.

\begin{figure*}[h]
\begin{tabular}{cc}
\hspace{-8pt} \includegraphics[scale=.26]{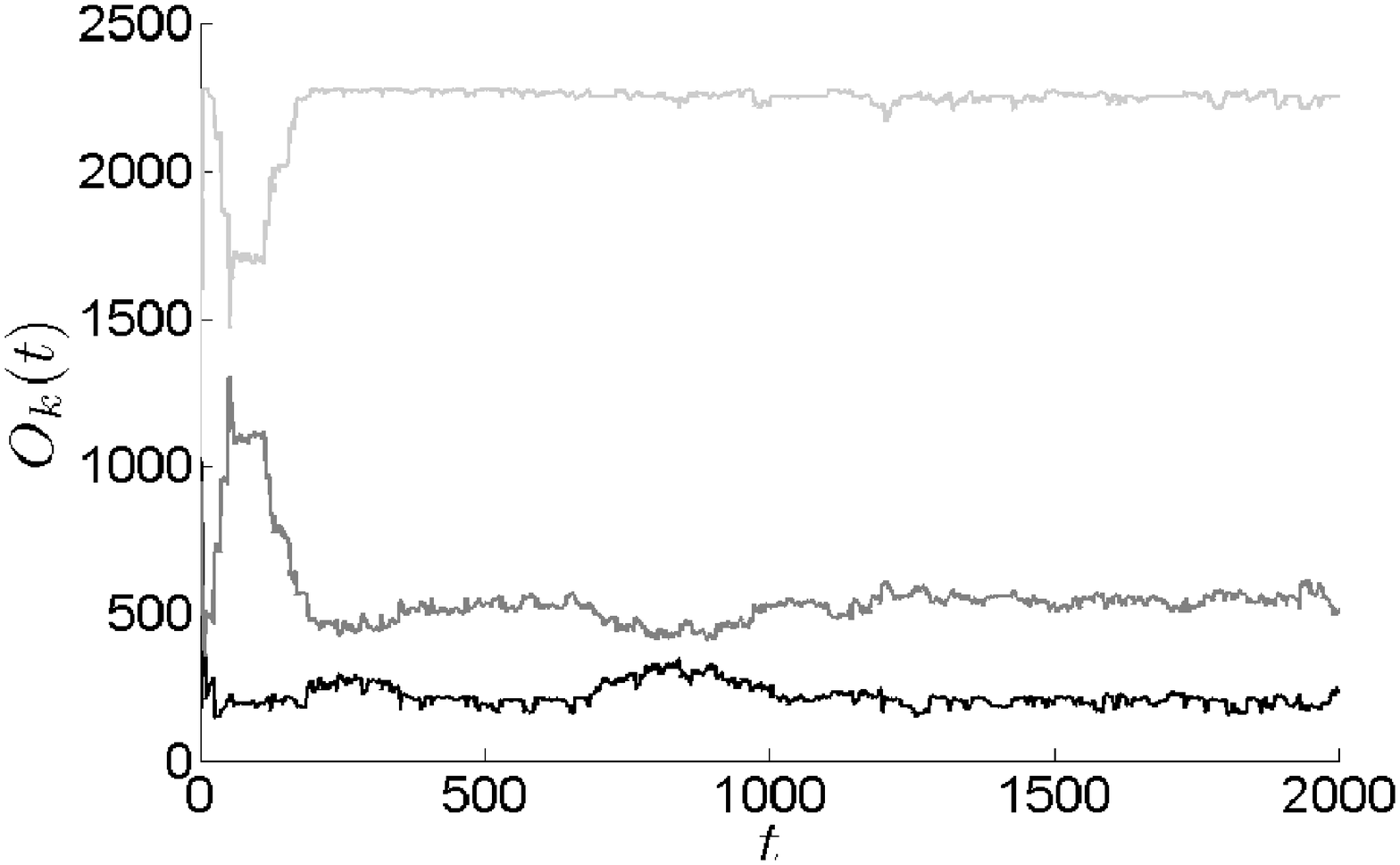} & \hspace{-8pt} \includegraphics[scale=.26]{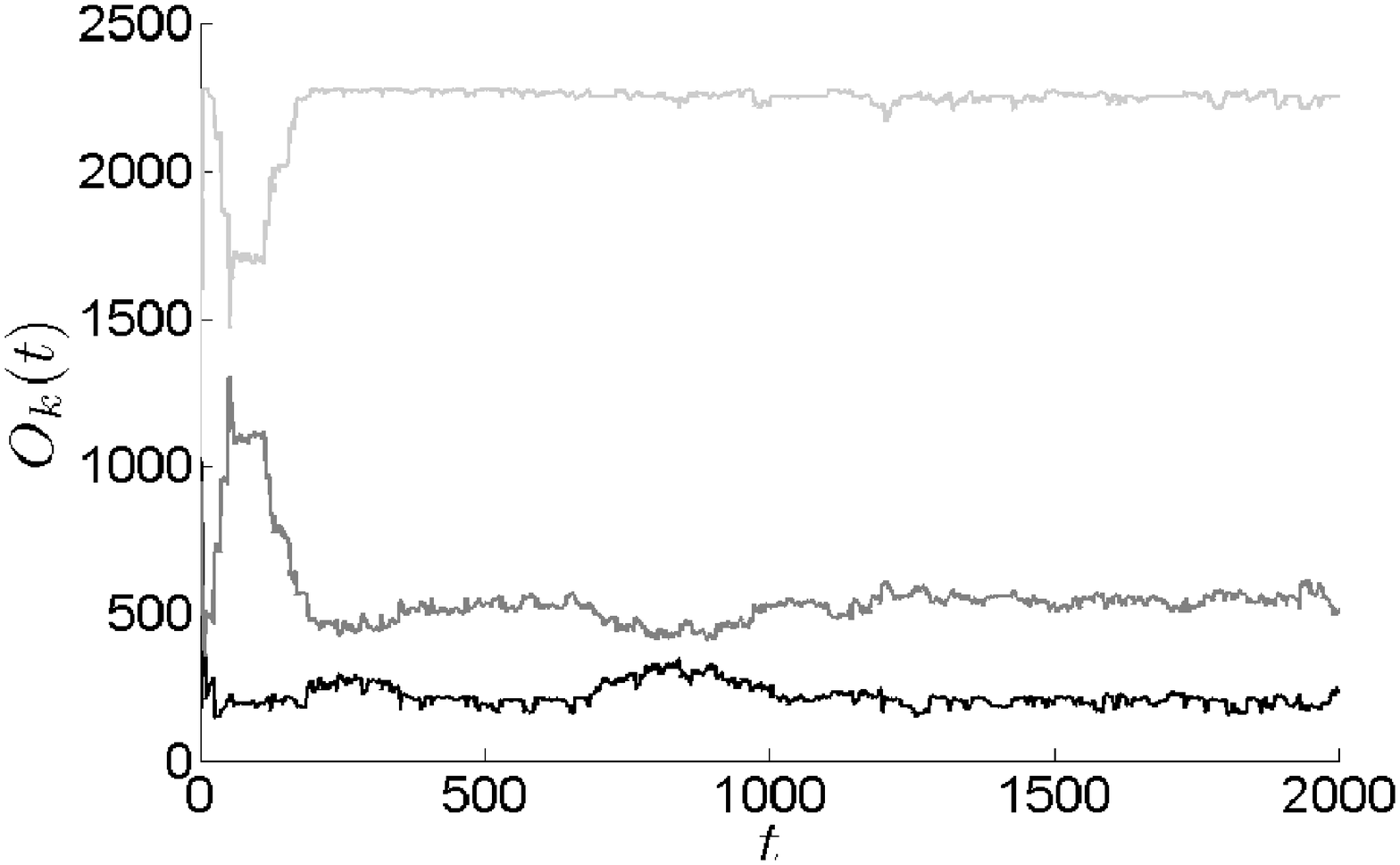} \\
\end{tabular}
\vspace*{8pt}
\caption{\label{fig010} Time evolution of market demands $A_k(t)$ (left) and occupancies $O_k(t)$ (right) for three regular markets with $N=3001$, $s=2$ and $m=5$. The lightest grey curves correspond to the largest market occupantion with the largest demand fluctuations.}
\end{figure*}

We also checked if the asymmetry is present for more than two markets of regular structure.
Considerations similar to discussions of Figs~\ref{fig3}, \ref{fig4} and \ref{fig5} lead us to the asymptotic theoretical formulae for market occupancies. 
If we order $K$ markets, beginning from the biggest one, we have
\begin{eqnarray}
\langle O_k\rangle = N \left \{ \begin{array}{lr} 1-1/2^s, & k=1 \\
                                   (1-1/2^s)(1/2^s)^{k-1}, & \quad\quad 1<k<K \\
                                   (1/2^s)^{K-1}. & k=K 
                 \end{array} \right .
\label{eq06}
\end{eqnarray}
We verified this result by simulations for three markets ($K=3$) and present market demands and occupancies in Figs~\ref{fig010}.
Similarly to the $K=2$ case, the occupancies stabilize and the largest occupancy corresponds to the market with the largest demand fluctuations.

\subsection{Irregular markets}

Consider irregular markets with variable $N_1$ and fixed $N_2=301$.
One expects that increasing $N_1$ amplifies the probability of high $A_1(t)$ and expects significant increase of utilities for all strategies on the first market.
Further reasoning follows that, for the regular MMG, high $A_1(t)$ has no impact on the market preference of agents playing on market 1 but seriously affects those on market 2.
The mean presence on market 1 grows indefinitely and is asymptotically linear with $N_1$
\begin{eqnarray}
\langle O_1\rangle \sim N_1+(1-1/2^s)N_2 \quad\quad\mbox{for}\quad N_1\rightarrow\infty.
\label{eq6}
\end{eqnarray}
For the mean presence on market 2 there exists the limit
\begin{eqnarray}
\lim_{N_1\rightarrow\infty} \langle O_2\rangle = N_2/2^s.
\label{eq7}
\end{eqnarray}
Fig.~\ref{fig8} presents simulations for mean occupancies and demand variances as functions of $Q=N_1/2^m$.
As seen in Fig.~\ref{fig8}, when the market 1 grows then the market 2 becomes more efficient.
This is understandable because if $N_2$ agents were separated then such market would work in the herd mode \cite{savit}.
Allowing agents to act also on market 1 spurs some of them to do that.
The larger $N_1$, the bigger fraction of $N_2$ prefers market 1 and the smaller fraction of $N_2$ prefers market 2.
Consequently, $\langle O_2\rangle$ converges to the limit (\ref{eq7}), equal to 75 for our simulation.
Decreasing $N_2$ makes smooth transition of market 2 from the herd mode to the cooperation mode, reflected in lower {\it per capita} fluctuations (cf. Fig.~\ref{fig8}).

\section{Conclusions and final remarks}

We considered the MMG as a natural generalization of the classical MG.
Although each agent has {\it a priori} the same probability to play on all markets, after some time most of them act on this market where relatively high aggregated demand occurs for the first time.
This spontaneous breakdown of choice symmetry depends on the payoff function $g$ in eqn (\ref{eq4}).
In particular, for $g(x)=x$ strong fluctuation of demand significantly affects utilities and the effect is well visible.
We checked that for weaker $g(x)$ the effect if less pronounced and for $g(x)=\sgn(x)$ it disappears.
This corroborates our conjecture that in the MMG there may exist markets strongly attracting clients.
How fast it occurs, it depends on the likelihood of strong demand fluctuation.
This reminds the well-known phenomenon on the scale-free networks where links are absorbed by nodes very inhomogenously and the initial random inhomogeneity of absorptive power of nodes determines their further imparity in course of the evolution.
This kind of behaviour can be also observed on real financial markets.
Large markets of high liquidity strongly attract agents because they enable easy conversion of chunks of assets into cash.
This constitutes a feedback mechanism making these markets even more liquid, etc.

In our model of the MMG we pointed out that the effect of symmetry breaking is visible only for $s\ge 2$.
However, the mathematical formalism introduced in the paper is true also for $s=1$.
As we mentioned in chapter 5, the market occupancies $O_k(t)$ stabilize at the levels $N/2^s$ and $N(1-1/2^s)$. 
Therefore the occupancies are equal for $s=1$ and the effect cannot be observed.
This explains why the authors of ref. \cite{bianconi}, using $s=1$, did not observe any symmetry breaking in their model.

We found that in the MMG there exists a critical value of $N/2^m$ above which agents behave asymmetrically.
In this mode we observe collectivity in their behaviour.
In particular, there always exists a history $\mu_k^C$ for which all agents on bigger market react identically.
We are aware of similarities of the asymmetric mode to some mechanisms on real financial markets.
The closest real analogy is perhaps financial crisis when majority of investors sell an asset.
However, there are important shortcomings of our model with respect to reality, e.g. restriction of collectiveness to given history $\mu_k^C$.
In addition, in reality individuals usually do not continue their game just after selling an asset but may stay apart for some time.
Therefore adequacy of our MMG to real markets is still limited.
We see some promising extensions of the classical MG which could improve the MMG.
Among them are the grand canonical approach \cite{slanina}, stochastic strategies \cite{hart} and categorization of investors~\cite{challet_4}.

\appendix

\section{The flow chart of the multi-market minority game simulation}

\begin{figure*}[h]
\begin{center}
\includegraphics[scale=.6]{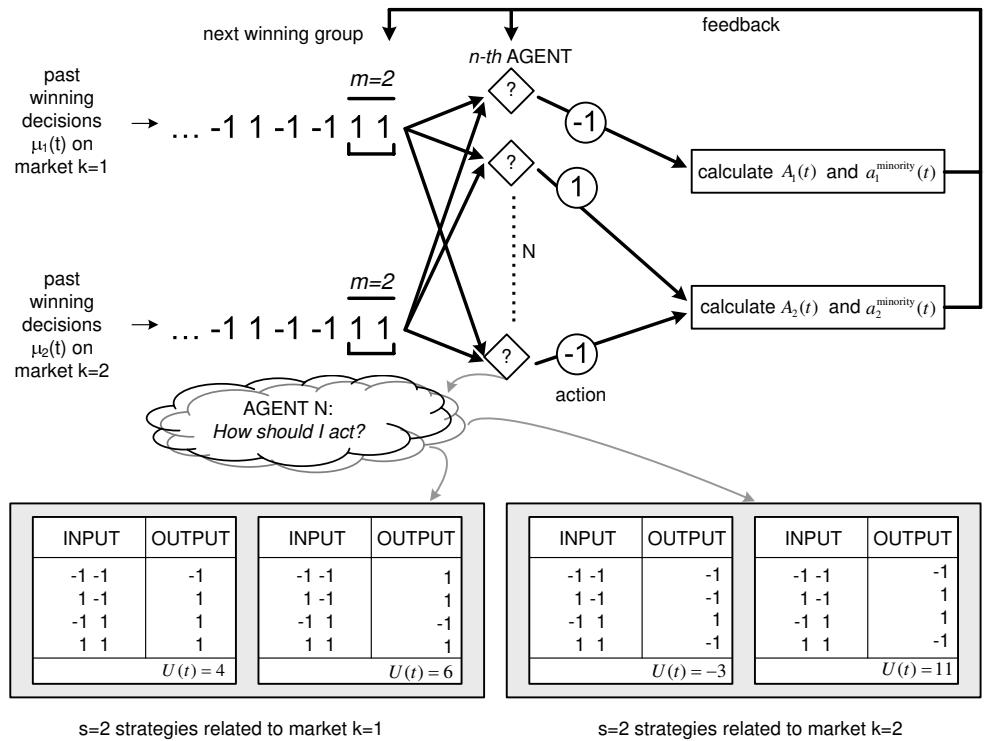} 
\end{center}
\vspace*{8pt}
\caption{\label{figapp} The diagram of the MMG simulation for $N$ agents and $m=2$ decision record length, $K=2$ markets and $s=2$ decisions for each agent and each market.}
\end{figure*}
The diagram in Fig.~\ref{figapp} presents how the MMG is simulated.
At any time step $t$ each of $N$ agents examines the public pattern $\mu_k(t)$ of $m$ most recent winning decisions on each of $K$ accessible markets ($k=1,\ldots,K$, where $K=2$ in our example in Fig.~\ref{figapp}).
Each agent acts according to its policy encoded in $s$ strategies per market ($s=2$ in Fig.~\ref{figapp}).
First, an agent chooses a strategy and then acts on the market assigned to this strategy.
At any time one agent acts on one market only.
Subsequently, the aggregate demand $A_k(t)$ and the minority action $a_k^\ast(t)$ on the $k$-th market are calculated according to the minority rule.
The $A_k(t)$ constitutes a feedback to agents, used by them in the adaptation process.
The $a_k^\ast(t)$ is then concatenated with $\mu_k(t)$ thus producing $\mu_k(t+1)$.
The patterns $\mu_k(t+1)$ are analysed by agents at $t+1$.

\end{document}